# Local Inversion Symmetry Breaking and Thermodynamic Evidence for Ferrimagnetism in Fe$_3$GaTe$_2$

Sang-Eon Lee[1,7], Yue Li[3], Yeonkyu Lee[4], W. Kice Brown[5], PeiYu Cai[6], Jinyoung Yun[4], Chanyoung Lee[4], Alex Moon[1,2], Lingrui Mei[1,2], Jaeyong Kim[7], Yan Xin[1], Julie A. Borchers[8], Thomas W. Heitmann[9,10,11], Matthias Frontzek[12], William D. Ratcliff[8,13,14], Gregory T. McCandless[5], Julia Y. Chan[5], Elton J. G. Santos[6,15,16], Jeehoon Kim[4], Charudatta M. Phatak[3], Vadym Kulichenko[1], Luis Balicas[*,1,2]

1. National High Magnetic Field Laboratory, Tallahassee, Florida 32310, United States

2. Department of Physics, Florida State University, Tallahassee, Florida 32306, United States

3. Materials Science Division, Argonne National Laboratory, Lemont, Illinois 60439, United States

4. Department of Physics, Pohang University of Science and Technology, Pohang 37673, Republic of Korea

5. Department of Chemistry & Biochemistry, Baylor University, Waco, Texas 76706, United States

6. Institute for Condensed Matter and Complex Systems, School of Physics and Astronomy, The University of Edinburgh, Edinburgh EH9 3FD, U.K.

7. Department of Physics, Hanyang University, Seoul 04763, Republic of Korea

8. NIST Center for Neutron Research National Institute of Standards and Technology Gaithersburg, MD 20899, USA.

9. University of Missouri Research Reactor, University of Missouri, Columbia, Missouri 65211, USA

10. Department of Physics and Astronomy, University of Missouri, Columbia, Missouri 65211, USA

11. MU Materials Science and Engineering Institute, University of Missouri, Columbia, Missouri 65211, USA

12. Neutron Scattering Division, Oak Ridge National Laboratory, Oak Ridge, Tennessee 37831, USA

13. Department of Physics, University of Maryland, College Park, Maryland 20742, United States

14. Department of Materials Science and Engineering, University of Maryland, College Park, Maryland 20742, United States

15. Higgs Centre for Theoretical Physics, The University of Edinburgh, Edinburgh EH9 3FD, U.K.

16. Donostia International Physics Center (DIPC), 20018 Donostia-San Sebastián, Basque Country, Spain

*balicas@magnet.fsu.edu

**KEYWORDS:** van der Waals ferromagnets, anomalous Hall-effect, skyrmions, topological Hall-effect, ferrimagnetism, transmission electron microscopy, magnetic force microscopy

**ABSTRACT**

The layered compound $Fe_3GaTe_2$ is attracting attention due to its high Curie temperature, low dimensionality, and the presence of topological spin textures above room temperature, making $Fe_3GaTe_2$ a good candidate for applications in spintronics. Here, we show, through transmission electron microscopy (TEM) techniques, that $Fe_3GaTe_2$ single crystals break local inversion symmetry while maintaining global inversion symmetry according to X-ray diffraction. Coupled to the observation of Néel skyrmions via Lorentz-TEM, our structural analysis provides a convincing explanation for their presence in centrosymmetric materials. Magnetization measurements as a function of the temperature displays a sharp first-order thermodynamic phase-transition leading to a reduction in the magnetic moment. This implies that the ground state of $Fe_3GaTe_2$ is globally ferrimagnetic and not a glassy magnetic state composed of ferrimagnetic, and ferromagnetic domains as previously claimed. Neutron diffraction studies indicate that the ferromagnetic to ferrimagnetic transition upon reducing the external magnetic field is associated with a change in the magnetic configuration/coupling between Fe1 and Fe2 moments. We observe a clear correlation between the hysteresis observed in both the skyrmion density and the magnetization of $Fe_3GaTe_2$. This indicates that its topological spin textures are affected by the development of ferrimagnetism upon cooling. Observation, via magnetic force microscopy, of magnetic bubbles at the magnetic phase boundary suggests skyrmions stabilized by the competition among magnetic phases and distinct exchange interactions. Our study provides an explanation for the observation of Néel skyrmions in centrosymmetric systems, while exposing a correlation between the distinct magnetic phases of $Fe_3GaTe_2$ and topological spin textures.

**TOC graphics**

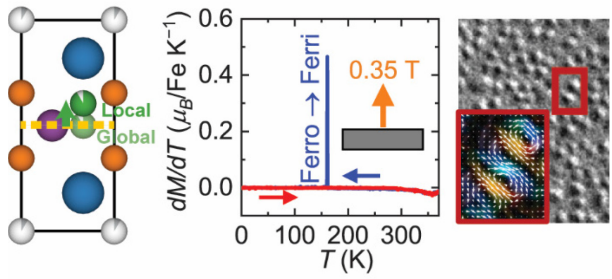

**Introduction**

Topological spin textures, such as merons, skyrmions, or hopfions, are one of the most active research areas in magnetism. Skyrmions were first observed in the helimagnet MnSi.[1,2] Its distinctive spin structure is topologically non-trivial, leading to its robustness.[3–5] This robustness has attracted considerable attention from the perspective of applications, e.g., as information carriers; therefore, understanding the mechanisms that stabilize and allow the manipulation of such topological spin textures is of technological relevance. Various techniques for manipulating skyrmions were developed, including the application of pulses of electrical currents,[6–8] light pulses,[4] and electrostatic gating.[9] Currently, it is anticipated that magnetic skyrmions might become a key element in cutting-edge technology, e.g., in neuromorphic[7,10–13] and quantum computing.[14–16]

The discovery of van der Waals (vdW) magnets, such as in $Cr_2Ge_2Te_6$[17] and $CrI_3$,[18] opened a new avenue in the study of magnetic materials, given their stability at the monolayer limit upon mechanical exfoliation. Substantial efforts have been devoted to finding magnetic skyrmions in vdW magnets for practical use, e.g., via large area growth of thin films for the development of magnetic skyrmion-based spintronics applications. Topological spin textures were found in several vdW magnets, including $Cr_2Ge_2Te_6$,[19,20] $Cr_xTe_2$,[21,22] $Fe_3GeTe_2$,[23–28] $Fe_5GeTe_2$,[29–32] $(Fe_{1-x}Co_x)_5GeTe_2$,[33,34] and $Fe_3GaTe_2$.[35–42] Among them, $Fe_3GaTe_2$ is regarded as an ideal platform for magnetic skyrmion-based applications due to its remarkably high Curie temperature, surpassing room temperature, or $T_c \geq 357$ K, and the various topological spin textures observed in this compound, such as Bloch-skyrmions,[36,37] Néel skyrmions,[35,39,40] hybrid-skyrmions,[37] and even the skyrmionium.[41] Notice that $Fe_3GaTe_2$ is already being actively pursued as the building block for room-temperature spin-valve devices.[43–46]

The variety of topological spin structures observed in $Fe_3GaTe_2$ suggests that several intertwined mechanisms lead to their stabilization. For instance, the Dzyaloshinskii-Moriya (DM) interaction, perhaps resulting from a disorder-induced global inversion symmetry-breaking structural transition towards the space group *P*3*m*1, from the original space group *P*6$_3$/*mmc*, was claimed to account for the observation of Néel-skyrmions in $Fe_3GaTe_2$.[35,36,40] On the other hand, the competition between the magnetic dipole and the DM interactions was claimed to stabilize Bloch- and hybrid-like skyrmions in $Fe_3GaTe_2$.[36,37] Lattice disorder and pinned domains were also claimed to be important factors contributing to the formation of topological spin textures in this compound.[41] Notice that it was recently shown that competing magnetic interactions leading to magnetic frustration, and their interplay with itinerant electrons via the Ruderman-Kittel-Kasuya-Yosida interaction could lead to magnetic skyrmions even in centrosymmetric materials.[47–51] Given the inherently correlated nature of magnetism, and the complexity introduced by disorder in real materials, it is challenging to pinpoint the exact mechanism leading to skyrmion formation. However, exposing the underlying mechanisms leading to skyrmion stabilization in a centrosymmetric, layered ferromagnetic compound like $Fe_3GaTe_2$ is important both from a fundamental perspective and for the development of applications.

Here, we demonstrate that the disorder observed in the structural position of the Fe2 ions within the unit cell of $Fe_3GaTe_2$, as observed via High-Angle Annular Dark-Field Scanning Transmission Electron Microscopy (HAADF-STEM), leads to an average Fe2 atomic location (when averaged over hundreds of atoms) that mildly breaks local inversion symmetry. In contrast to previous studies,[35,36,40] where disorder-induced inversion symmetry suppression leads to global inversion symmetry breaking, we found, through single crystal X-ray diffraction (SCXRD), that the crystal structure of $Fe_3GaTe_2$ is well modeled by a centrosymmetric space group. This might explain reports claiming to observe Néel skyrmions in $Fe_3GaTe_2$, despite clear global inversion symmetry according to X-ray diffraction.[41,52] From magnetization measurements, we observed a sharp, first-order-like, spin-flip transition under field cooling conditions from a metastable FM single-domain phase to a FiM ground state. The FiM state did not return to the FM state under field warming conditions, pointing to the metastable nature of the FM state, which leads to a pronounced thermal hysteresis in the magnetization. Neutron diffraction measurements reveal that the FM to FiM transition results from a change in magnetic coupling between Fe1 and Fe2 atoms.

The coexistence between interlayer ferromagnetic and antiferromagnetic couplings was assumed and labeled as ferrimagnetism in a previous surface sensitive study[41] of $Fe_3GaTe_2$. Density functional theory (DFT) calculations claimed that interlayer antiferromagnetism is the ground state of $Fe_3GeTe_2$.[53] However, our neutron scattering study is consistent with an intralayer ferrimagnetic phase triggered by a change in the magnetic coupling between Fe1 and Fe2 atoms. This should compel further theoretical investigations.

Based on the concrete evidence for the coexisting among magnetic phases, we found that the topological spin textures, i.e, skyrmions in $Fe_3GaTe_2$, could result from the coexistence among magnetic structures or equivalently competition among exchange couplings. Our MFM images reveal the possible presence of topological spin textures at field values near the magnetic phase boundary between FiM and FM states. This indicates that the competition between FM and FiM states might play a role in the formation of topological spin textures in $Fe_3GaTe_2$. Lorentz transmission electron microscopy (LTEM) images reveal that the Néel skyrmion density displays thermal hysteresis which is concomitant with the thermal hysteresis observed in the magnetization implying that the skyrmion density is affected by the development of the FiM ground state upon cooling. This suggests that the temperature degree of freedom could be used to manipulate skyrmion density and size.

**Results and Discussion**

**Structural characterization of $Fe_3GaTe_2$**

Single crystals of $Fe_3GaTe_2$ were prepared via chemical vapor transport (CVT) and characterized using single-crystal X-ray diffraction (SCXRD) (see Supporting Figure S1 for precession images and Supporting Tables S1, S2, and S3 for details on the crystallographic data, refinement parameters, and atomic coordinates). The diffraction data were best fitted to the space group $P6_3/mmc$, with unit cell dimensions $a$ = 4.0793(3) Å, and $c$ = 16.107(3) Å as provided in Table S1, which are consistent with a previous report.[54] Figure 1a depicts the crystallographic structure of $Fe_3GaTe_2$. A notable aspect is the Fe deficiency at the Fe2 site, whose occupancy is within the 88 - 90% range for all five crystals studied. Most of the missing Fe2 atoms end up intercalated between the vdW gaps originating the Fe3 sites whose occupancy falls within the range 8.7 - 9.0% range, while still yielding the approximate ratio Fe:Ga:Te = 3:1:2 (see Supporting Table S3). Energy dispersive X-ray spectroscopy also yields a stoichiometric ratio Fe:Ga:Te = 3:1:2 within its standard resolution (see Supporting Figure S2 and Table S4). This type of disorder was observed in previous studies of $Fe_3GaTe_2$[36,41] and in its isostructural compounds $Ni_3Ge(Ga)Te_2$[55–57] as well as in $Fe_3GeTe_2$.[58] Fe intercalation tends to increase the Curie temperature for both $Fe_3GeTe_2$[58] and $Fe_3GaTe_2$.[36,41]

The atomic structure of $Fe_3GaTe_2$ (Figure 1a) was investigated via HAADF-STEM (Figure 1b and 1c). For the statistical analysis, we conducted Gaussian fittings on 864 of atoms observed in the image (see, Supporting Figure S3 for the fitted image and the resulting atomic coordinates for the analyzed region). Figures 1c and 1d are magnified views of Figure 1b, without and with the coordinate markers resulting from the Gaussian fittings of the atomic intensities, respectively. We found that the coordinates of the Fe2 and Ga ions are offset vertically with respect to the mirror plane (indicated by the dashed yellow line) situated between top and bottom Te atoms within the unit cell, generating local disorder. Their average vertical offset is shown in Figure 1e, which displays a unit cell resulting from the average atomic coordinates. We found a vertical offset $\Delta y = 0.16 \pm 0.06$ Å and $0.17 \pm 0.07$ Å for the upper and lower Fe2 atoms, respectively, with $\Delta y = 0.05 \pm 0.08$ Å and $0.04 \pm 0.07$ Å for the upper and lower Ga atoms, respectively, where the numbers after the plus/minus signs are the standard deviations (see, Supporting Figure S4 for probability distributions of atomic displacements with respect to the mirror plane between Te atoms). These vertical offsets of the Fe2 ions lead to a mildly broken local inversion symmetry. In previous SCXRD studies, vertical offsets of 0.0871 Å (Ref. [35]) and 0.076(1) Å in (Ref. [40]) were reported for Fe2, which not only agree with this analysis of the HAADF-STEM images, but are also within its standard deviation. In Ref. [36], a substantial vertical offset was also reported for the Ga atoms, i.e., approaching 0.29(4) Å on average, based on SCXRD. In sharp contrast to our study, all three references, claim that $Fe_3GaTe_2$ crystallizes in the non-inversion symmetric $P3m1$ space group instead of the originally proposed inversion symmetric $P6_3/mmc$. This points to a higher level of disorder in the samples studied in those reports. In our samples, the vertical offset is clearly observed, but only in the HAADF-STEM images. However, SCXRD refinement in multiple crystals clearly yields the $P6_3/mmc$ space group for the crystals synthesized for this study. This agrees with the original report on this compound implying global structural inversion symmetry, despite the TEM analysis. We conclude that the vertical displacements of the Fe2 ions would break inversion symmetry over relatively short distances, likely forming domains with distinct average values for the Fe2 displacements, while preserving global inversion symmetry. A complete characterization of these structural domains and their influence on Néel skyrmions formation will be the subject of a future study. This conclusion contrasts with the results and conclusions in Refs. [35,36,40] claiming a distinct, non-inversion symmetric space group on a global scale. A possible path to reconcile this apparent discrepancy is to assume that crystals grown via distinct synthesis protocols are prone to distinct amounts of disorder. With disorder leading, in some crystals, to a global lack of inversion symmetry that contrasts with the single crystals synthesized by us.

The very mild signal of the intercalated Fe3 atoms can also be identified in our HAADF-STEM images (see Supporting Figure S5).

**Magnetic properties of $Fe_3GaTe_2$**

In the following paragraphs, we discuss magnetization and Hall effect measurements performed on the same crystal, although 19 crystals were characterized. A detailed account of the dependence of the electrical transport and magnetization data on crystal quality can be found in Ref. [59]. Here, we chose the crystal of the highest quality by selecting the one displaying the highest longitudinal and Hall conductivities among the 19 crystals measured. To investigate its magnetic properties, we measured the magnetization of $Fe_3GaTe_2$ over a wide range of temperatures ($T$) and magnetic fields ($\mu_0 H$). In Figures 2a and 2b we plot the magnetic susceptibility $\chi = M/(\mu_0 H)$ as a function of $T$ under $\mu_0 H = 250$ Oe oriented along the $c$-axis and $ab$-plane of the crystal, respectively. The much larger size of $\chi$ for $\mu_0 H$ oriented along the $c$-axis (20.5 memu/g·Oe at $T = 2$ K) when compared to the value (1.38 memu/g·Oe at $T = 2$ K) extracted for fields along the $ab$-plane, reveals a large perpendicular magnetic anisotropy. By differentiating the ZFC curves in Figures 2a and 2b with respect to temperature, we estimate the ferromagnetic transition temperature to be around 360 K, as shown in Figure S6, which is consistent with the neutron diffraction measurements (see Supplementary Figure S7). Additional structures can also be observed in the derivative at low temperatures (Figure S6). In addition, neutron scattering measurements performed in zero field support a ground state with magnetic moments pointed collinearly along the *c*-axis (see Supplementary Figure S8). The quantitative analysis of magnetic anisotropy was extracted via magnetization measurements as a function of magnetic field along the *ab*-plane and Hall effect measurements with a small angle $\theta = 1.7°$ between the direction of the field and that of the applied electrical current (see Supplementary Figure S8). From these measurements one extracts a value ~0.40 meV per Fe atom ($1.7 \times 10^6$ J/m$^3$) at $T = 2$ K and ~0.19 meV per Fe atom ($7.9 \times 10^5$ J/m$^3$) at $T = 300$ K, which is larger than the previously reported value of $4.79 \times 10^5$ J/m$^3$ at $T = 300$ K.[54]

Among field-cooled and field-warmed traces, a clear thermal hysteresis loop is observed for magnetic fields oriented along the *c*-axis (Figure 2a), being pronounced in the temperature range of ~ 200 K ≤ $T$ ≤ ~ 350 K. This hysteresis loop is observed in a region of temperatures where a dip displaying a minimum at $T = 290$ K is observed in the ZFC susceptibility. A similar dip is also observed for fields along the *ab*-plane (Figure 2b). These anomalies and the concomitant thermal hysteresis loop suggest the possibility of an additional phase transition upon cooling from room $T$, perhaps associated with the development of ferrimagnetism (FiM), which will be discussed below.

The thermal hysteresis becomes more pronounced in the susceptibility data collected under higher magnetic fields. Figure 2c shows $\chi$ traces as a function of $T$ under $\mu_0 H = 0.35$ T. Notably, one observes a sharp, step-like decrease in the magnetization at $T = 161$ K upon field cooling, which is also clearly shown in the derivative with respect to temperature (Figure S6). This obvious FiM state does not return to the FM state upon field warming up to $T_c$, pointing to the metastable nature of the FM state. Notice that it is also observed in the temperature-dependent anomalous Hall conductivity (see Supporting Figure S9). On the other hand, the magnetic transition does not affect the resistivity, implying that the magnetic transition does not induce a concomitant change in the carrier density or mobility (see Supporting Figure S9). Given that the ground state becomes FiM, we conclude that the FM phase observed above the transition is metastable and stabilized by the Zeeman energy induced by the external field. This implies that FiM and FM are nearly degenerate in energy, with the perturbation provided by a small magnetic field being enough to stabilize a metastable FM state. We extracted the phase diagram by measuring the susceptibility as a function of the temperature under field cooling conditions for various magnetic fields, i.e., for 0.2 T ≤ $\mu_0H$ ≤ 0.375 T, (Figure 2d). We observed that the FM to FiM transition temperature decreases rapidly upon increasing the magnetic field, i.e., from $T$ = 341 K under $\mu_0H$ = 0.2 T to $T$ = 99 K under $\mu_0H$ = 0.37 T (see Figure S6 for the derivatives with respect to temperature). Above $\mu_0H$ = 0.375 T, we could no longer detect the transition.

We also studied the magnetic properties via the field dependence of the magnetization at distinct temperatures (Figure 2e). Ferromagnetic behavior accompanied by a hysteresis loop was observed all the way up to $T$ = 350 K. The magnetization saturates at a value of 1.97 $\mu_B$/Fe at $T$ = 2 K under $\mu_0H$ = 7 T. Notably, a sudden abrupt decrease or increase in the magnetization is observed when the magnetization leaves the spin-polarized state. Such asymmetric magnetization displaying a step upon decreasing the magnetic field is observed in many systems and attributed to domain formation once the system is brought back from full saturation, see, e.g., Ref. [60]. However, this sharp change also resembles the FM to FiM transition observed in the $\chi$ as a function of the $T$ curves. To expose the transition more clearly, we defined the difference in $M$, $\Delta M$, by subtracting the magnetization measured with an increasing magnetic field from the one measured with a decreasing field (see, inset in Figure 2e). As seen in Figure 2f, which shows $\Delta M$ as a function of $\mu_0H$ for various temperatures, $\Delta M$ displays maxima at the fields where the transition occurs.

We compared the transition observed in the temperature-dependent measurements (under fixed field, or FF) with the transition observed in field-dependent measurements (fixed temperature, or FT). We obtained pairs of $T_{FF}$ and $\mu_0 H_{FF}$ from the sudden drop of the magnetic susceptibility in Figure 2d (e.g., as indicated by the black arrow) and pairs of $T_{FT}$ and $\mu_0 H_{FT}$ from the peaks observed at positive fields in Figure 2f (e.g., as indicated by the blue arrow). $\mu_0 H_{FF}$ as a function of $T_{FF}$ and $\mu_0 H_{FT}$ as a function of $T_{FT}$ are plotted in Figure 2g. Overall, both plots coincide, indicating the same origin for both data sets. Since the FiM state is accompanied by domain formation, we interpreted the transition as a transition from the spin-polarized, single FM-domain state into a mixed (FM + FiM) multi-domain one. Therefore, $\mu_0 H_{FF}$ as a function of $T_{FF}$ and $\mu_0 H_{FT}$ as a function $T_{FT}$ plots depict the phase boundary between FiM and FM phases. Remarkably, in the lower temperature regime ($\leq$ 25 K), the transition field decreases upon decreasing the temperature, which contrasts with the high-temperature regime ($\geq$ 25 K). This explains the absence of the transition above $\mu_0 H$ = 0.375 T in the field-cooling measurement.

To explain the metastable nature of the FM state, we introduced a simple mean-field model accounting for the mechanism triggering the transition to the FiM phase from the FM single-domain state. We assumed that there are two distinctive magnetic moments, $m_1$ and $m_2$, where $m_1$ is fixed along the c-axis and $m_2$ could couple antiferromagnetically to $m_1$. Within the mean-field approach, we introduced an exchange field, $H_{ex}$, to determine the energy difference between the FM and FiM states. This leads to a total effective magnetic field of $H_{eff} = H_{ext} + H_D + H_{ex}$, where $H_{ext}$ is the external field, and $H_D$ is the demagnetization field. With the magnetic anisotropy $K$, the effective energy of $m_2$ is given by $E = -\mu_0 m_2 H_{eff} \sin\phi + K\cos^2\phi$, where $\phi$ is the angle between the ab-plane and the direction of $m_2$. Figure 2h shows the energy of $m_2$ as a function of angle. The magnetic anisotropy $K$ creates an energy barrier between FM and FiM states, and the positive (negative) $H_{eff}$ favors FM (FiM). The FM to FiM transition can be understood as the size of a negative $\mu_0 m_2 H_{eff}$ exceeding $2K$, resulting in an unstable FM state from the metastable FM state. We investigated the values of $H_D$ and $H_{ext}$ at the FM to FiM transition. We found that the size of $H_D$ is always larger than the size of $H_{ext}$ at the transition (see Supporting Figure S10).

The nature of the magnetic thermal hysteresis observed in Figure 2a can be more carefully investigated via AC magnetic susceptibility measurements. Figures 2i and 2j show the in-phase and out-of-phase components of the AC magnetic susceptibility respectively, both collected under distinct excitation frequencies. For both components, we did not observe any notable frequency dependence for frequencies ranging from 0.17 to 47.7 Hz. Furthermore, we observe no difference between the AC magnetic susceptibility measurements collected upon cooling and warming (Figures 2k and 2l). These results indicate that the thermal hysteresis shown in Figure 2a is not related to glassiness or slow spin dynamics. The absence of evidence for glassiness is difficult to reconcile with a previous claim for a disorder-induced spin freezing transition in $Fe_3GaTe_2$.[41]

The sudden drop in the magnetization observed in the field sweeps was claimed to occur through a simple spin reversal process that leads to the formation of labyrinthine domains.[60–62] It turns out that this mechanism is unphysical for $Fe_3GaTe_2$, given its large anisotropy field of $2K/M_s \approx 7$ T and the small demagnetization field of $\approx -0.6$ T at $T = 2$ K. In $Fe_3GaTe_2$, the external field required to overcome the energy barrier generated by the magnetic anisotropy and produce spin reversal is $\mu_0 H \approx -6.4$ T, when the actual magnetization step occurs at $\mu_0 H \approx 0.3$ T at $T = 2$ K. This large discrepancy implies a magnetic phase transition as the origin of the magnetization step instead of the formation of labyrinthine domains.

**Analysis of the neutron scattering data collected under applied magnetic fields**

To capture the difference between the magnetic structures pertaining to the FM and FiM regimes, we conducted neutron diffraction measurements under $\mu_0H = 1$ T (FM state) and $\mu_0H = 0.35$ T (FM + FiM mixed domains) at $T = 100$ K. The measurement under $\mu_0H = 0.35$ T was conducted after field cooling from room temperature to $T = 100$ K with the field fixed at this value. The measurement under $\mu_0H = 1$ T was collected by increasing the field from $\mu_0H = 0.35$ T to $\mu_0H = 1$ T at a fixed temperature right after collecting data under $\mu_0H = 0.35$ T. We subtracted the neutron diffraction pattern collected under $\mu_0H = 0.35$ T from the one collected at $\mu_0H = 1$ T to focus on their difference in intensities due to the respective magnetic arrangements. Figures 3a, 3b, and 3c display the diffraction patterns collected in $\mu_0H = 1$ T, $\mu_0H = 0.35$ T, and their subtraction, respectively. We found that the intensities of the (100), (1-10), and (0-10) peaks increase as the magnetic fields decrease from $\mu_0H = 1$ T to $\mu_0H = 0.35$ T.

To understand the increase in peak intensities in a smaller magnetic field, we simulated the neutron scattering intensity. The neutron scattering intensity can be expressed as a summation of the square of the structure factors from both nuclear and magnetic contributions: $|F(\mathbf{q})|^2 = |F_N(\mathbf{q})|^2 + |F_M(\mathbf{q})|^2$. The nuclear structure factor, $F_N(\mathbf{q})$, and magnetic one, $\mathbf{F}_M(\mathbf{q})$, can be expressed respectively as:

$$F_N(\mathbf{q}) = A \sum_j e^{-\frac{1}{2}\mathbf{q}^T U_j \mathbf{q}} b_j e^{-2\pi \mathbf{q} \cdot \mathbf{x}_j}, \tag{1}$$

$$\mathbf{F}_M(\mathbf{q}) = A\gamma r_0 \frac{1}{2} g \sum_j \left[ \hat{\mathbf{m}}_j - \hat{\mathbf{q}}(\hat{\mathbf{m}}_j \cdot \hat{\mathbf{q}}) \right] e^{-\frac{1}{2}\mathbf{q}^T U_j \mathbf{q}} f_j(q) S_j e^{-2\pi \mathbf{q} \cdot \mathbf{x}_j}, \tag{2}$$

where $A$ is a scaling factor, $\mathbf{q}$ is the scattering vector, U is the Debye-Waller matrix factor, $b$ is the nuclear scattering length, $\mathbf{x}$ is the atomic position vector, $\gamma$ is the gyromagnetic ratio for neutrons, $r_0$ is the classical electron radius, $g$ is the Landé $g$-factor, $\hat{\mathbf{m}}$ is the unit vector of the magnetic moment, $f(q)$ is the magnetic form factor, and $S$ is the spin quantum number. The index $j$ runs through all atoms in the unit cell. Equation (2) is derived for the $Q = 0$ case, corresponding to all magnetic peaks superimposed onto nuclear ones. Experimentally, we did not observe any $Q \neq 0$ peaks.

Figure 3d shows the simulated peak intensities from a unit cell ($A = 1$) for the observed ($hkl$) peaks. We simulated three cases: a nonmagnetic state, a FM state (parallel Fe1 and Fe2 moments), and a FiM state (antiparallel Fe1 and Fe2 moments). For these simulations we used the structural parameters obtained from the X-ray diffraction data (the elements of the U matrix at $T = 100$ K were approximated to be one-third of those at room temperature). We found that the differences among (0-10), (1-10), and (100) peak intensities are most pronounced between the FM and FiM states among all observed peaks and that the peak intensities for the FiM case are significantly larger than those for the FM case. This is consistent with the proposed scenario of a polarized FM state under $\mu_0 H = 1$ T, but a FiM state under $\mu_0 H = 0.35$ T.

Assuming that the Fe1 and Fe2 moments maintained the same alignment (or the same magnetic moment ratio $m_{Fe1}/m_{Fe2}$), the neutron diffraction intensity measured under $\mu_0 H = 1$ T should increase by 29% with respect to the one collected under $\mu_0 H = 0.35$ T for (0-10), (1-10), and (100) peaks given the 1.8 $\mu_B$ per Fe measured at $\mu_0 H = 1$ T (see Figure 2e and Figure 3e for the magnified view) versus the 1.38 $\mu_B$ per Fe at $\mu_0 H = 0.35$ T (see Figure 2c and Figure 3f for the magnified view) in the magnetization. In the calculations based on Eqs. 1 and 2, we assume the moments of Fe1 and Fe2 to be equal, but the Fe3 one to be zero. Instead, when the magnetic field increases from $\mu_0 H = 0.35$ T to $\mu_0 H = 1$ T, the experimentally measured intensities for the (0-10), (1-10), and (100) peaks decrease by 17, 20, and 16 %, respectively. This indicates realignment between moments [see, respective intensities for the (1-10) peak in Figure 3g]. In our simulations we assume $m_{Fe1} = m_{Fe2} = 2~\mu_B$. With these values, the simulated intensities for the (0-10), (1-10), and (100) peaks should decrease by 46% when the system is driven from the FiM to the spin-polarized or FM state.

**The anomalous Hall conductivity of Fe$_3$GaTe$_2$**

Next, we focus on the electrical transport properties of Fe$_3$GaTe$_2$ and on their relationship with the magnetic phases and domain coexistence. It is already known that Fe$_3$GaTe$_2$ displays the anomalous Hall effect which follows its magnetization[35,52,54,59,63–65] hence, we measured the Hall conductivity for our crystals. Figure 4a displays the Hall conductivity $\sigma_{xy}$ for a Fe$_3$GaTe$_2$ single crystal collected at several $T$s. The value of $\sigma_{xy}$ was evaluated through the relation of $\sigma_{xy} = \rho_{yx}/(\rho_{yx}^2 + \rho_{xx}^2)$ (see the raw data of $\rho_{yx}$ and $\rho_{xx}$ in Supporting Figure S11). The shape of the Hall conductivity mimics the field dependence of the magnetization (Figure 2e), indicates that the anomalous Hall conductivity dominates the overall Hall response. The large value of $\sigma_{xy} \approx 420$ $\Omega^{-1}$cm$^{-1}$ suggests that the anomalous Hall effect is governed by the intrinsic Berry curvature mediated mechanism.[66–69] By comparing $\sigma_{xy}(\mu_0 H, T)$ to the magnetization data, we found evidence for the FM to FiM transition also in the Hall response. Figure 4b shows $\sigma_{xy}$ as a function $\mu_0 H$ at $T$ = 4.25 K within the low magnetic field regime (−1 T ≤ $\mu_0 H$ ≤ 1 T), revealing the detailed structure of the Hall signal. Similarly to what was observed for the magnetization, an abrupt decrease or increase in $\sigma_{xy}$ is observed when $\sigma_{xy}$ decrements from its saturated value upon decreasing $\mu_0 H$. We defined the difference in $\sigma_{xy}$, $\Delta\sigma_{xy}$, by subtracting $\sigma_{xy}$ with increasing magnetic field from the $\sigma_{xy}$ trace collected with decreasing magnetic field. Figure 4c shows $\Delta\sigma_{xy}$ as a function of $\mu_0 H$ at various temperatures. As for the magnetization, the FM to FiM transition produces peak like structures in the plot of $\Delta\sigma_{xy}$ as function of $\mu_0 H$.

**Magnetic domains in Fe$_3$GaTe$_2$ and their evolution as a function of magnetic field**

Here, we delved into the magnetic properties of Fe$_3$GaTe$_2$, focusing on the evolution of the magnetic domains as a function of the magnetic field via MFM imaging. Figures 4d to 4g display Magnetic Force Microscopy (MFM) images measured under magnetic fields $\mu_0 H$ = 0.35, 0.2, −0.2, and −0.4 T, respectively. In Figure 4b each magnetic field value is indicated by dots superimposed onto the anomalous Hall conductivity curves to indicate their relative proximity to saturation. These measurements were conducted upon decreasing magnetic fields from $\mu_0 H$ = 0.35 T to $\mu_0 H$ = −0.4 T at $T$ = 4.25 K. At $\mu_0 H$ = 0.35 T, the image reveals the absence of magnetic domain structure, which is consistent with the saturating magnetization shown in Figure 4d. Once the magnetic field is decreased to $\mu_0 H$ = 0.2 T, i.e., upon crossing the FM to FiM boundary (Figure 4e), we observe three separate magnetic regimes depicted by blue, dark blue, and bright yellow regions. The narrow, bright yellow regions can be interpreted as spin-reversed regions with respect to the magnetic field. The large blue region can be interpreted as a ferromagnetic background, as in Figure 4d. We found that dark blue regions are evidence for a sudden increase in magnetic susceptibility, which is consistent with a phase boundary between FM and FiM states (see Supporting Figure S12 and related notes). These observations indicate that the single-domain FM state has transitioned into a coexisting FM and FiM multi-domain state rather than a single FiM domain. With further decreasing the magnetic field to $\mu_0 H$ = −0.2 T, the magnetic domain structure becomes more complex, forming intricate labyrinthine-like domains (Figure 4f). With further increasing the field strength ($\mu_0 H$ = −0.4 T), some of the labyrinthine domains transform into magnetic bubbles (dark blue circles in Figure 4g), likely skyrmions superimposed onto a blue FM background. These magnetic bubbles formed close to the phase boundary (Figure 4g) suggest that their development is favored by competing interactions and the metastable conditions.

Under $\mu_0 H = -0.4$ T, the Zeeman-effect and the FiM coupling compete against each other, albeit the Zeeman effect remains slightly more dominant, leading to a state with a predominance of FM domains. It is noteworthy that the bubble structure only appears when the system is entering into the FM state rather than leaving the FM state. This indicates that the formation of magnetic bubbles requires the initial presence of the labyrinthine-like domain structure.

**LTEM images unveil Néel skyrmions in Fe$_3$GaTe$_2$**

We investigated the presence of topological spin textures via Lorentz TEM, or LTEM imaging. An exfoliated crystal of Fe$_3$GaTe$_2$ with a thickness $t \cong 45$-$60$ nm, transferred onto a 50 nm thick silicon nitride grid under inert conditions, was used for imaging. Figure 5a shows the experimental configuration used for the LTEM. We fixed the magnetic field at $\mu_0 H = 250$ Oe during image collection under field cooling conditions (see Supporting Figure S13 for the LTEM images under different field values) while varying the tilt angle α between the sample and the magnetic field. Figures 5(b-d) display the LTEM images collected at $T = 120$ K for α = $-25$, $-2.7$, and $+25°$, respectively, after field cooling from $T = 373$ K. The LTEM contrast of bubble-like spin textures can only be recognized at higher tilt angles but disappear without tilting, indicating that they correspond to Néel skyrmions.[37] The existence of Néel skyrmions is also confirmed by micromagnetic simulations and reconstructed LTEM images with the proper choice of the DM coupling (see, Methods and Figure S14). This observation is consistent with previous studies reporting Néel-type skyrmions in Fe$_3$GaTe$_2$.[35,36,40,41,52] Although we must mention that were also able to simulate Bloch-like spin textures in hybrid skyrmions by increasing the saturating magnetization. This is consistent with other studies claiming Bloch or hybrid skyrmions stabilized by the dipole-dipole interaction (see, Methods and Figure S14).[36,37] Importantly, our observations imply that Néel-type skyrmions can be stabilized by local inversion symmetry breaking, even in the presence of global inversion symmetry. To the best of our knowledge, this has not yet been demonstrated. This provides a convincing answer to the puzzle of Néel-type skyrmions in an inversion symmetric system as implied by our XRD measurements.[41,52]

Subsequently, we investigated the formation and density of Néel-skyrmions as a function of the temperature. Figures 5(e-h) show LTEM images collected at $T = 323$ K, 283 K, 183 K, and 123 K with a fixed angle $\alpha = -21°$, after cooling under field from $T = 373$ K. The skyrmion density sharply decreases upon field cooling from $T = 323$ K to $T = 183$ K, while the average skyrmion diameter increases from ~80 nm to ~115 nm. On the other hand, the density and size remain nearly constant when the field is cooled from $T = 183$ K to $T = 123$ K. In contrast, upon field-warming, the skyrmion density remains almost constant from $T = 123$ K all the way up to $T = 283$ K as shown in Figures 5(h-j) but increases above 283 K from which value the individual skyrmions begin to merge into labyrinthine domains to the point of becoming unrecognizable at $T = 323$ K (Figure 5k). Note that the previous study also reported a denser skyrmion density of $Fe_3GaTe_2$ at higher temperatures[39]. This behavior is clearly exposed in Supporting Movies 1 and 2. Our LTEM images (Figure 5) include some merged or elongated magnetic bubbles in addition to Néel skyrmions, which were not observed in Ref. [35] whose samples break global inversion symmetry. These elongated structures, observed in our samples, might indicate modulations in local structural disorder locally weakening the Dzyaloshinskii–Moriya interaction and hence suppressing Néel skyrmion formation. Conversely, samples that depart from the ideal stoichiometry ought to be characterized by a higher level of disorder, which is likely to lead to broken global inversion symmetry favoring the DMI interaction and thus skyrmion formation.

The distinct behavior displayed by the skyrmion density when compared to their radii upon field cooling and warming exposes their thermal hysteric behavior. One can compare this hysteretic behavior with the thermal hysteresis observed in magnetization. For this purpose, the skyrmion density, obtained by hand counting their number per unit area in the LTEM images, is plotted in Figure 5l along with the hysteresis in the magnetization δ$M$. δ$M$ is evaluated by subtracting the magnetization measured on field warming from the magnetization measured on field cooling under $\mu_0 H$ = 250 Oe. It turns out that one observes a clear correlation between both thermal hysteretic behaviors. Note that the thermal hysteresis loop observed in $Fe_3GaTe_2$ has been understood as resulting from competing magnetic interactions.[70] Considering our result, the thermal hysteresis loop observed in Figure 2a could be understood as resulting from a competition between FM and FiM, as a broad version of the first-order transition observed in Figure 2d. Therefore, this clearly indicates a correlation between competing FM and FiM interactions and skyrmion density.

A similar thermal hysteresis for skyrmions was reported for $Fe_3GeTe_2$ which displays a similar FM to FiM transition in the field-dependence of its magnetization.[71] This study showed that skyrmion diameters increased with decreasing temperature, locally forming closely packed skyrmions. These closely packed skyrmions merged with nearby skyrmions, thus lowering the skyrmion density upon reducing $T$. In our study, we did not observe closely packed skyrmions while the skyrmion radii increased once $T$ was reduced. The observed correlation between skyrmion density and the FM to FiM transition would suggest that their density decreases once one crosses the FM to FiM phase boundary upon reducing the temperature. This thermal hysteresis, likely observed at the FM to FiM phase boundary, would explain why the skyrmion density does not recover its original value after increasing the temperature towards room $T$. Since the FM and FiM regions are indistinguishable in the LTEM images, this technique cannot provide the exact distribution of their respective domains.

Néel-type skyrmions stabilized by a non-zero DMI interaction was explained in terms of the asymmetric interaction among neighboring Fe (Fe1 and Fe2) mediated by the Te atoms and attributed to the vertical displacement of the Fe2 atoms with respect to the mirror plane between Te.[35,40] Here, we observed a similar vertical displacement of the Fe2 atoms in stoichiometric $Fe_3GaTe_2$ as well as Néel-type skyrmions. Notice that Fe deficiency is claimed to play a crucial role in breaking the inversion symmetry in isomorphic $Fe_3GeTe_2$.[72] However, vertical displacement of the Fe2 atoms does not lead to global inversion symmetry breaking in our stoichiometric compound. We speculate that Fe3 intercalation precludes the lack of global inversion symmetry. This is supported by the observation of global inversion symmetry coexisting with Néel-type skyrmions in another study on Fe-intercalated $Fe_3GaTe_2$.[41]

An important aspect to be understood is the role of the relative size between structural domains and topological spin textures. If the skyrmion diameter spanned several domains characterized by negative and positive average displacements of the Fe2 atoms relative to their equilibrium position, one would expect a diminished role for the DMI interaction. This might explain why $Fe_3GaTe_2$ tends to host Bloch- or hybrid-like skyrmions as the DMI decreases[36,37] or why in Figure 5 we observe elongated spin bubbles coexisting with Néel skyrmions.

**Unconventional Hall effect in $Fe_3GaTe_2$**

In this section, we analyze both the Hall effect and the magnetization of $Fe_3GaTe_2$ to extract a possible topological Hall signal, which would be evidence for topological spin textures. It was shown theoretically that the Hall resistivity can be subjected to three contributions, i.e., $\rho_{yx} = \rho^O + \rho^A + \rho^T$, where $\rho^O$ is the ordinary Hall resistivity, $\rho^A$ the anomalous Hall resistivity, and $\rho^T$ the topological Hall resistivity.[73,74] Here, we will utilize the term unconventional Hall resistivity for the remnant Hall resistivity (once $\rho^O + \rho^A$ are subtracted) and labeled as $\rho^U$, instead of the topological Hall resistivity term, since it would require an unambiguous confirmation of its topological origin. $\rho^A$ is proportional to $M$, which can be expressed as $R_S M$, where $R_S$ is the anomalous Hall coefficient.[75] We evaluated $R_S$ by scaling $\rho_{yx}$ with respect to the saturating magnetization at the lowest magnetic fields. This eliminates the contribution of $\rho^O$ superimposed onto the saturated region of $\rho_{yx}$. $\rho^O$ is evaluated by fitting $\rho_{yx} - \rho^A$ to a smooth function, although a proper evaluation of $\rho^O$ is difficult due to its very small size relative to the other Hall components. Figure 6a plots $\rho_{yx}$, $\rho^A$, and $\rho^O$ as functions of the magnetic field.

Within the low field regime $\rho^O$ is negligible, with this allowing us to directly compare $\rho_{yx}$ to $\rho^A$ and $\rho^U$. Figure 6b shows a comparison between $\rho_{yx}$ and $\rho^A$ as a function of the magnetic field within the low field regime ($0 \leq \mu_0 H < 0.65$ T). A discrepancy between $\rho_{yx}$ and $\rho^A$ becomes pronounced when the magnetization enters (or leaves) the FM state, as indicated by the magenta (navy) ellipses. This discrepancy becomes more pronounced when the magnetization suddenly changes upon leaving the FM as $\mu_0 H$ is reduced. Figure 6c displays $\rho^U$ as a function of $\mu_0 H$ where $\rho^U$ is obtained by subtracting $\rho^O + \rho^A$ from $\rho_{yx}$. When the magnetization enters the FM state, the shape of the peaks observed in $\rho^U$ is relatively broad and smooth, while they sharpen considerably as the magnetization leaves the FM state. These spike-like peaks are likely attributed to a slight difference in the actual value of the magnetic field between magnetization and transport measurements, whose effect could be pronounced when $dM/d(\mu_0 H)$ is large (sharp change). See Supporting Figure S15 for non-spike containing unconventional Hall resistivities as the result of the smoother transitions observed in lower quality samples. On the contrary, the smooth peaks observed in $\rho^U$ when $\mu_0 H$ increases are perhaps evidence for an anomaly, or an unconventional Hall response associated to the transition from the FiM to the spin polarized state.

To make this point clearer, in Figure 6d we plot the position of the peaks extracted from the smooth unconventional Hall resistivities (see, Figure S16 for the extraction of the peak positions) $\mu_0 H_U$, along with the temperature and magnetic field values defining the magnetic phase boundary. In a wide range of temperatures of 25 K ≤ $T$ ≤ 300 K, $\mu_0 H_U$ lies at the phase boundary between FM and FiM states, thus suggesting that it may correspond to an anomaly due to the phase transition between both phases. The exact coincidence between $\mu_0 H_U$ and the magnetic phase boundary is observed in the other two lower quality samples, although superposition becomes less evident as the sample quality decreases (see Figure S15). However, below $T$ = 25 K, $\mu_0 H_{FT}$ decreases upon decreasing temperature, while $\mu_0 H_U$ increases. It turns out that one observes an upturn in the resistivity upon cooling below $T$ = 22 K (see Figure S15), suggesting that the bifurcation in the phase boundary is associated with this upturn, previously claimed to result from Kondo-like correlations.[68] We also plot in Figure 6d the experimental paths followed for the MFM and LTEM measurements shown in Figure 4(d-g) and Figure 5(e-h), respectively. This exposes two possible ways to generate skrymions in Fe$_3$GaTe$_2$: i) field cooling from above $T_c$ under a small magnetic field, and ii) approaching the saturation of magnetization or the magnetic phase boundary from the labyrinthine domain phase, as reported in Ref. [39].

One might also associate the peak observed in unconventional Hall resistivity with a topological Hall contribution, given that the smooth unconventional Hall effect shown in Figure 6c peaks when the magnetization is about to reach saturation. This interpretation would be consistent with the emergence of magnetic bubbles in the MFM images (see blue circle in Figure 6d) or the reported presence of skyrmions when the sample is subjected to a field sweep.[39] However, smooth peaks also occur when the Fe$_3$GaTe$_2$ is driven away from magnetic saturation or even at zero field, particularly in the lower-quality samples (Figure S15). Furthermore, the magnitude of the putative topological Hall peak scales with the anomalous Hall resistivity of the sample. At $T = 2$ K, $\rho^U$ ranges from 0.12 $\mu\Omega$ cm in the smallest resistivity sample ($\rho_{xx} \approx 96.4$ $\mu\Omega$ cm and $\rho_{yx} \approx 3.99$ $\mu\Omega$ cm under $\mu_0 H = 1$ T, leading to $\rho^U/\rho_{yx} \sim 3\%$) to 0.83 $\mu\Omega$ cm in the largest resistivity one ($\rho_{xx} \approx 521$ $\mu\Omega$ cm and $\rho_{yx} \approx 18.6$ $\mu\Omega$ cm under $\mu_0 H = 1$ T yielding $\rho^U/\rho_{yx} \sim 4.5\%$). Given that the topological Hall resistivity is expected to follow $\rho_{xy}^T = \gamma n_{sk}\, e^2/h$, with $\gamma$ being a factor determined by the electronic band structure, $h$ the Planck constant, and $e$ the electron charge,[74] one would be unable to explain this scaling of the putative topological Hall signal with the resistivity of the distinct crystals. A possible origin for the unconventional Hall resistivity might be the difference in $R_s$ values between ferrimagnetic and ferromagnetic regimes. It has already been pointed out that a spatially inhomogeneous $R_s$ could lead to a topological-Hall-like signal,[76] and this could well be the case for $Fe_3GaTe_2$ given the coexistence of domains of distinct magnetic phases.

Figure 6e shows $R_s$ as a function of the temperature upon cooling and warming and exposes a clear difference in $R_s$ values before and after the magnetic transition between the spin polarized and the FM + FiM multidomain regimes. $R_s$ changes by ~3.3% upon crossing the magnetic phase boundary, which is comparable to the ratio $\rho^U/\rho_{yx} \sim 3\%$ at $T = 2$ K. This scenario would explain the increase in unconventional Hall resistivity as resistivity increases. Given the negligible change in resistivity at the magnetic field-induced phase transition (see Figure S9), the difference in $R_s$ ought to result from the difference in $S_H = \sigma_{xy}/M$, which is directly related to the Berry curvature[75] (see Figure S17 for the difference in $S_H$ between both magnetic phases). A change in Berry curvature between FM and FiM phases might be caused by the difference in exchange fields or the electron hopping integral between Fe1 and Fe2 atoms. A modification in electron-magnon couplings associated with the different magnetic states could also contribute to the different $S_H$ values as an extrinsic mechanism.[75] However, the current data does not allow us to unambiguously identify the origin of the unconventional Hall effect. Hence, we cannot exclude the possibility of a topological origin associated with spin textures.

One study claimed a surprisingly large topological Hall resistivity maximum of 8 µΩ cm in $Fe_3GaTe_2$ at 10 K[35]. This value would be much larger than so-called giant topological Hall responses reported for $Mn_2PtSn$ that peaks at ~ 1.5 µΩ cm at $T$ = 150 K,[77] $NdMn_2Ge_2$ (1.35 µΩ cm at $T$ = 300 K)[78] or $Fe_3Sn_2$ (2.01 µΩ cm at room temperature).[79] However, it is not evident how the magnetization from the exfoliated $Fe_3GaTe_2$ crystal was evaluated to compare it with the Hall response. In contrast, we extracted a putative topological Hall effect from a bulk single crystal by measuring both quantities, the Hall effect and the magnetization, on the same crystal. Another study claimed to observe a large and unconventional topological Hall effect in $Fe_3GaTe_2$ at 250 K, i.e., peaking a 1.48 µΩ cm when the magnetic field is applied along the *ab*-plane.[80] However, we can reproduce very similar Hall signals only when the magnetic field is slightly misaligned away from the *ab*-plane, with no resolvable Hall signal when the magnetic field is exactly aligned along the *ab*-plane (see Figure S18). A clear and large topological Hall signal of 5.4 µΩ cm at 10 K was also claimed to have been observed in (oxidized $Fe_3GaTe_2$)/(pristine $Fe_3GaTe_2$) heterostructures.[81] However, the peak occurs when the system is brought away from, and not upon entering magnetic saturation. Given the broad range of reported values and behaviors, the presence of a topological Hall response in $Fe_3GaTe_2$ remains as a matter of further studies.

**Conclusions**

Through a detailed structural analysis via transmission electron microscopy, we show that the average positions of certain atoms, i.e., Fe2, can lead to broken inversion symmetry in $Fe_3GaTe_2$ over a characteristic length scale even in the presence of global inversion symmetry according to X-ray diffraction. We observed Néel-type skyrmions via Lorentz-TEM, demonstrating that it is possible to stabilize Néel-type skyrmions when inversion symmetry is broken at the local level. We highlight a sharp magnetic transition from a single metastable FM domain to a multi-domain state (FM + FiM) in the temperature and field dependences of both the magnetization and the anomalous Hall effect. The presence of ferrimagnetism is consistent with the analysis of our neutron scattering measurements. Our simulations of the neutron diffraction data indicates that the magnetic transition results from a change in the magnetic configuration/coupling between Fe1 and Fe2 atoms. The connection between the coexistence of magnetic phases and the observation of topological spin textures, is indicated by the observation of magnetic bubbles, likely skyrmions, at the phase boundary between FiM and FM. This connection is further supported by the correlation between the thermal hysteresis observed in the magnetization, and the temperature-dependence of the skyrmion density. Our study suggests that the competition between magnetic phases, in this case FM and FiM might contribute to the stabilization of topological spin textures. Although this competition can lead to magnetic frustration, we do not detect any evidence for glassiness or spin freezing, despite previous claims for both $Fe_3GaTe_2$ and its Ge analog. Glassiness is likely to be sample-dependent, i.e., dependent on the amount of disorder present in a particular crystal. Furthermore, global inversion symmetry in $Fe_3GaTe_2$ (space group $P6_3/mmc$ as determined in the present study and also in Refs. [41,54]) or its absence (space group $P3m1$ as proposed by other groups[35,36,40]), is likely dependent upon the amount of disorder (e.g., average global displacement of the Fe2 atom) induced by the synthesis protocol used. Such disorder is likely to favor distinct spin textures (even at the local level and in the same crystal), such as the Néel skyrmions reported here, and the hybrid skyrmions reported elsewhere. Finally, the anomaly observed in the Hall effect at the boundary between the FiM phase and the spin polarized state, previously attributed to a topological Hall component, could result from inhomogeneous $R_s$ or $S_H$ resulting from the coexisting of magnetic phases characterized by different textures and net values of Berry curvature.

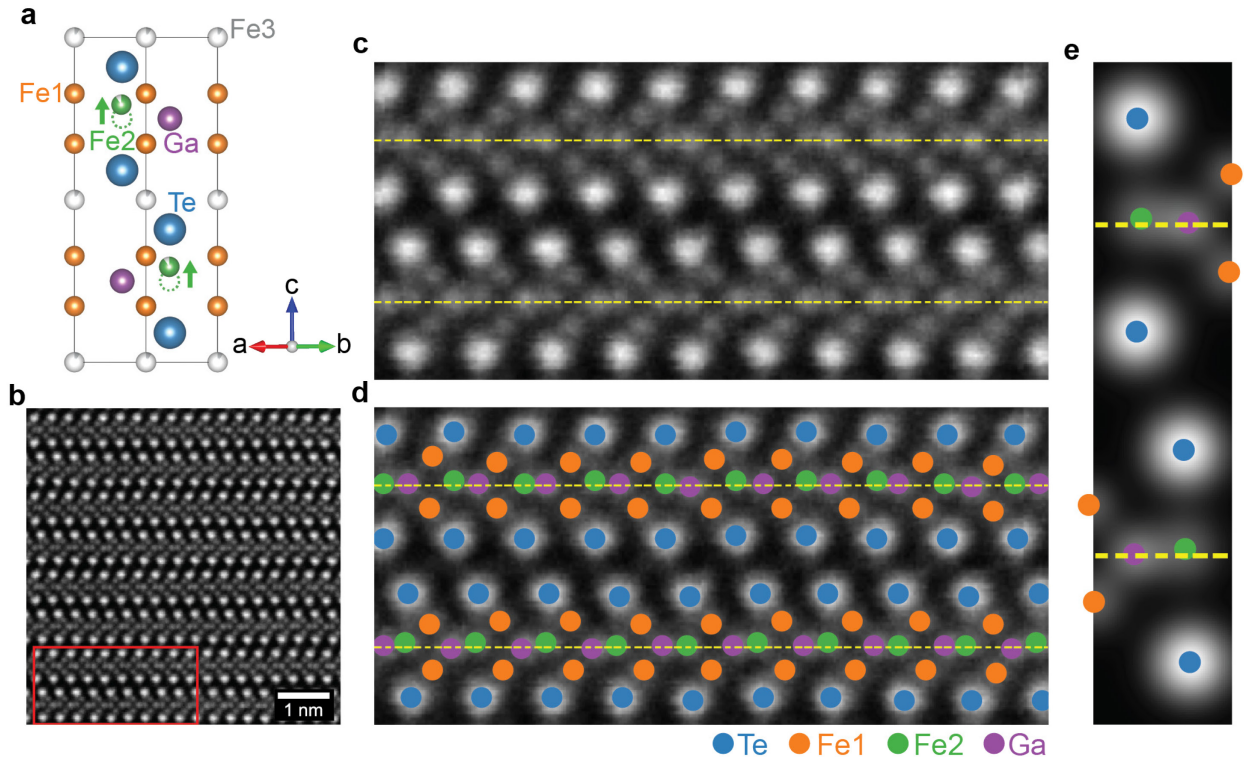

**Figure 1. Structural analysis of Fe$_3$GaTe$_2$. a,** Unit cell of Fe$_3$GaTe$_2$ as determined through single crystal X-ray diffraction (SCXRD) with the concomitant lattice parameters $a$ = 4.0793(3) Å and $c$ = 16.107(3) Å. The refinement confirmed the intercalation of Fe (8.8%) at the interstitial Fe3 site. Green dashed circles and green solid balls represent the global position of the Fe2 atoms, as determined by SCXRD, and the locally displaced average position of the same Fe2 atom observed via High-Angle Annular Dark-Field Scanning Transmission Electron Microscopy (HAADF-STEM) imaging, respectively. **b,** HAADF-STEM image of Fe$_3$GaTe$_2$ viewed from (110) direction. **c and d,** Magnified HAADF-STEM images with and without position markers. The magnified region is indicated by the red box in **b**. Position markers indicate the estimated coordinates for the atomic centers obtained from Gaussian fittings. It is shown that the Fe$_3$Ga chains are encapsulated by Te atoms forming a monolayer, and each layer is van der Waals coupled to nearby layers. The
29

vertical Fe2 and Ga positions vary throughout the unit cells, with the Fe2 positions having a vertical offset with respect to the mirror plane (indicated by yellow dashed lines) located between both Te atoms (blue markers). **e**, Unit cell of $Fe_3GaTe_2$ obtained by averaging the atomic coordinates via a Gaussian fitting of their HAADF-STEM atomic intensities. The vertical displacement for Fe2 with respect to the mirror plane are $\Delta y = 0.16 \pm 0.06$ and $0.17 \pm 0.07$ Å for the upper and lower atoms, respectively. The extracted displacements for the Ga atoms are $\Delta y = 0.05 \pm 0.08$ and $0.04 \pm 0.07$ Å for the upper and lower atoms, respectively. Here, the numbers after the plus/minus signs are the standard deviations.

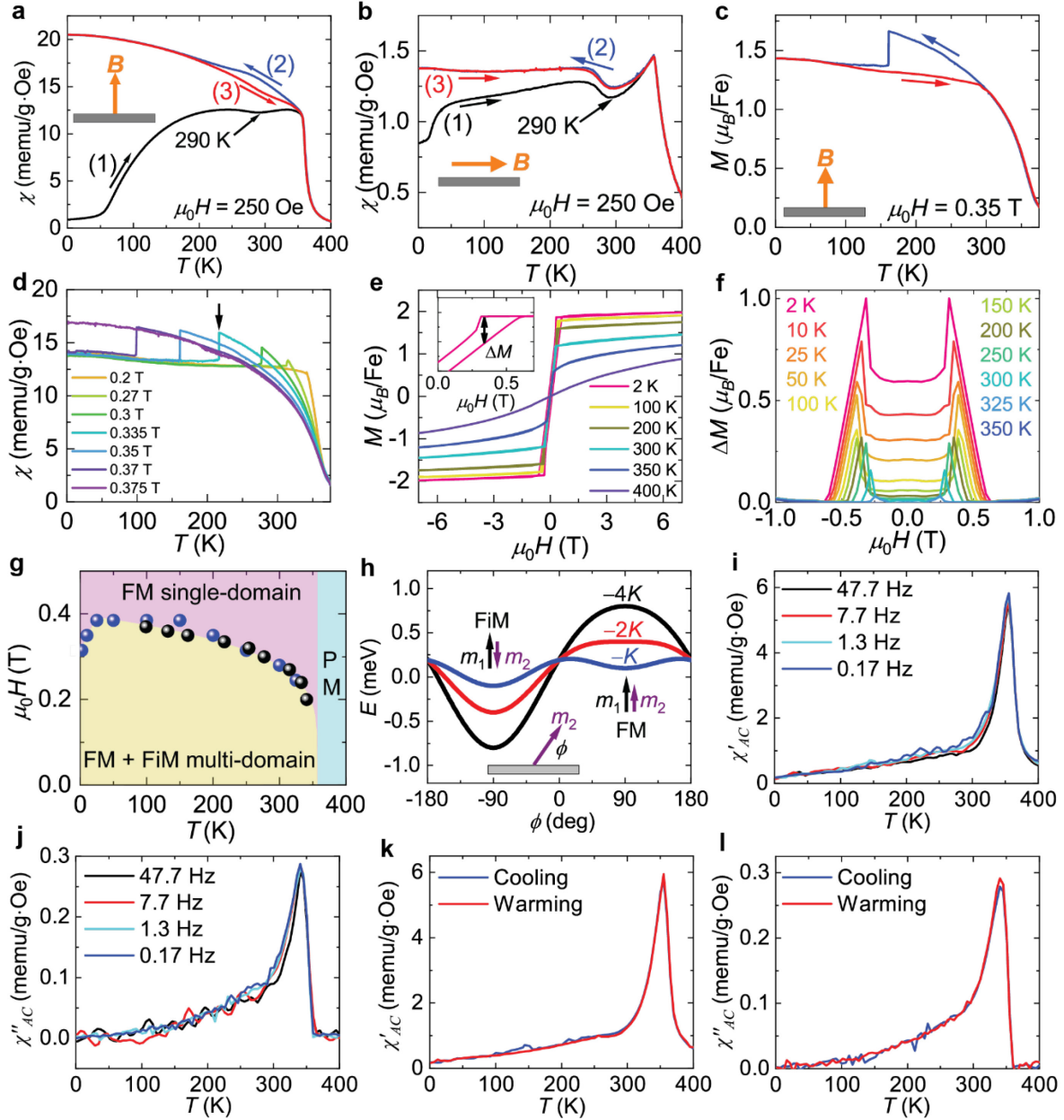

**Figure 2. Magnetic properties of Fe$_3$GaTe$_2$ and its magnetic phase diagram. a and b,** Magnetic DC susceptibility as a function of the temperature $T$ under $\mu_0 H$ = 250 Oe applied along the $c$-axis and $ab$-plane of crystal, respectively. Black, blue, and red solid lines display the zero-field cooled, field-cooled, and field-warmed data, respectively. The sequence of temperature-dependent measurements is indicated by the numbers in parentheses. In the zero-field cooled trace, dip

structures appear around $T = 290$ K, implying ferrimagnetic or antiferromagnetic coupling. **c,** Magnetization $M$ as a function of $T$ under a magnetic field $\mu_0 H = 0.35$ T applied along the $c$-axis. Blue and red solid lines display the field-cooled and field-warmed data, respectively. A sharp drop is observed in $M$ at $T = 161$ K, which is irreversible, leading to a pronounced thermal hysteresis loop. **d,** Magnetic susceptibility $\chi$ as a function of $T$ upon field cooling. Black arrow indicates a first-order phase transition under $\mu_0 H = 0.335$ T. Notice that a similar step in the magnetization as a function of $T$ was also observed in the ferromagnets $Tm_2Fe_{14}B$ and $Er_2Fe_{14}B$, and similarly interpreted as a spin reorientation transition.[82,83] **e,** $M$ as a function of $\mu_0 H$ indicating that FM is present at $T$s exceeding 350 K. Inset: Magnified view of the field-dependent $M$ in the low field regime, exposing its hysteresis. The difference in magnetization, $\Delta M$, is defined as the difference between $M$ measured under decreasing and increasing fields. Interval between major ticks along the $y$-axis is 1 $\mu_B$/Fe. **f,** $\Delta M$ as a function of $\mu_0 H$ at fixed $T$s. Blue arrow points to one of the peaks at $T = 2$ K. **g,** Magnetic field as a function of the temperature magnetic phase-diagram indicating, FM, ferrimagnetic (FiM), and paramagnetic (PM) phases. FM single-domain regime is colored in magenta, the coexistence regime between FM and FiM in yellow, and the PM above $T = 357$ K in blue. **h,** Angle-dependent effective energy of the magnetic moment $m_2$. We used $E = -\mu_0 m_2 H_{eff} \sin\phi + K\cos^2\phi$, where $H_{eff}$ is the effective magnetic field, and $K$ is the magnetic anisotropy. The inset shows the configuration of measurements. Black, red, and blue lines correspond to $\mu_0 m_2 H_{eff} = -4K$, $-2K$, and $-K$, respectively. Metastable FM state transits into the FiM state at $\mu_0 m_2 H_{eff} = -2K$. **i, j, k, and l,** AC magnetic susceptibility, real ($\chi_{AC}'$) and imaginary ($\chi_{AC}''$) components as functions of $T$. These traces were collected under a magnetic field $\mu_0 H = 250$ Oe with an AC excitation $\mu_0 H_{exc} = 5$ Oe. **i** and **j,** Respectively, $\chi_{AC}'$ and $\chi_{AC}''$ components of the magnetic AC susceptibility for different AC frequencies. These traces were collected upon decreasing the temperature. **k** and **l,** $\chi_{AC}'$ and $\chi_{AC}''$ components of the AC susceptibility measured with a frequency $f = 1.3$ Hz upon cooling and warming, respectively. Notice the absence of hysteresis.

**a** $\mu_0 H = 1$ T

**b** $\mu_0 H = 0.35$ T

**c** *Diff*

**d** nonmag / FM / FiM

**e** 1.8 $\mu_B$ at $\mu_0 H = 1$ T, $T = 100$ K

**f** $\mu_0 H = 0.35$ T, 1.38 $\mu_B$ at $T = 100$ K

**g** $\mu_0 H = 0.35$ T, $\mu_0 H = 1$ T

**Figure 3. Analysis of the neutron scattering data collected under magnetic fields applied along the *c*-axis. a, b, and c,** Neutron diffraction patterns collected under $\mu_0 H = 1$ T, and $\mu_0 H = 0.35$ T, and their subtraction, respectively. **d,** Simulated scattering intensities from a unit cell for the experimentally observed (*hkl*) peaks. Simulation is based on Eqs. (1 and 2) with 2 $\mu_B$ for both Fe1 and Fe2 atoms when simulating either the FM or the FiM state. For simplicity, the magnetic moments of the Fe3 atoms were fixed to $m_{Fe3} = 0$ $\mu_B$. **e,** Magnified view of Figure 2e for the $T = 100$ K data. **f,** Magnified view of Figure 2c for a field cooling trace under $\mu_0 H = 0.35$ T. **g,** Integrated intensities of the (1-10) peaks measured under $\mu_0 H = 1$ T (black marker) and under $\mu_0 H = 0.35$ T (red marker), illustrating the increase in magnetic signal upon reducing the magnetic field. This is consistent with our simulations, pointing to spin realignment once the magnetization saturates, or a transition from FiM to a FM state as implied by the temperature dependence of the magnetization shown in Figures 2c and 2d. However, a reduction in the average magnetic moments of either Fe1 or Fe2 (< 1 $\mu_B$) would also be compatible with this analysis.

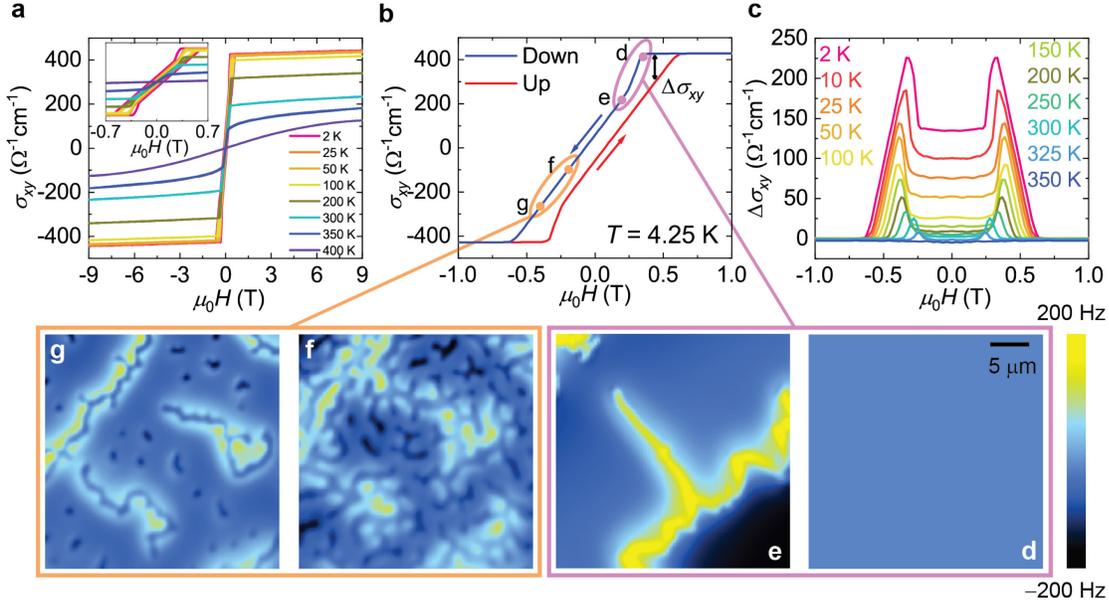

**Figure 4. Anomalous Hall conductivity and the magnetic domain structure of Fe$_3$GaTe$_2$. a,** Hall conductivity $\sigma_{xy}$ as a function of the magnetic field for several $T$s. $\sigma_{xy}$ as a function of $\mu_0 H$ follows the behavior of $M$ as a function $\mu_0 H$ (Figure 2e), indicating that the anomalous Hall conductivity is the dominant component of $\sigma_{xy}$. Inset: Magnified view of the field-dependence of the magnetization in the low field regime, exposing its hysteresis. **b,** Hall conductivity at $T = 4.25$ K in a narrow low field range. Net hysteresis in the Hall conductivity, $\Delta\sigma_{xy}$, defined as the difference between $\sigma_{xy}$ measured upon decreasing magnetic fields and the one collected on increasing field. **c,** Plot of $\Delta\sigma_{xy}$ as a function of $\mu_0 H$, displaying peaks at the FM to FiM transition. Similar structure is observed also in magnetization. **d, e, f, and g,** MFM images collected at $T = 4.25$ K under $\mu_0 H = 0.35, 0.2, -0.2,$ and $-0.4$ T (values indicated by the dots in panel **b**), respectively. These measurements started from $\mu_0 H = 0.35$ T, and then the field was progressively lowered to $-0.4$ T. At the magnetic field corresponding to the onset of the saturation of $\sigma_{xy}$ at $\mu_0 H = 0.35$ T, one observes no domain structure, which is compatible with a fully saturated magnetization (panel **d**). When the magnetic field is reduced to $\mu_0 H = 0.2$ T, which is just below

the field value required to saturate the magnetization, large labyrinthine-like domains start to emerge (panel **e**). These larger labyrinthine-like domains change to smaller and more complex ones by adjusting the field value to $\mu_0 H = -0.2$ T (panel **f**). Here, blue and yellow colors represent magnetic domains with their moments pointing along and antiparallel to the magnetic field, respectively. By further decreasing the magnetic field to $\mu_0 H = -0.4$ T, which is close to the magnetic field value required for saturation, bubble-like domains, i.e., perhaps skyrmions, are observed. MFM cannot expose the texture of the in-plane magnetic moments in each domain, therefore, with this technique it is not possible to determine their character, i.e., topological, or not.

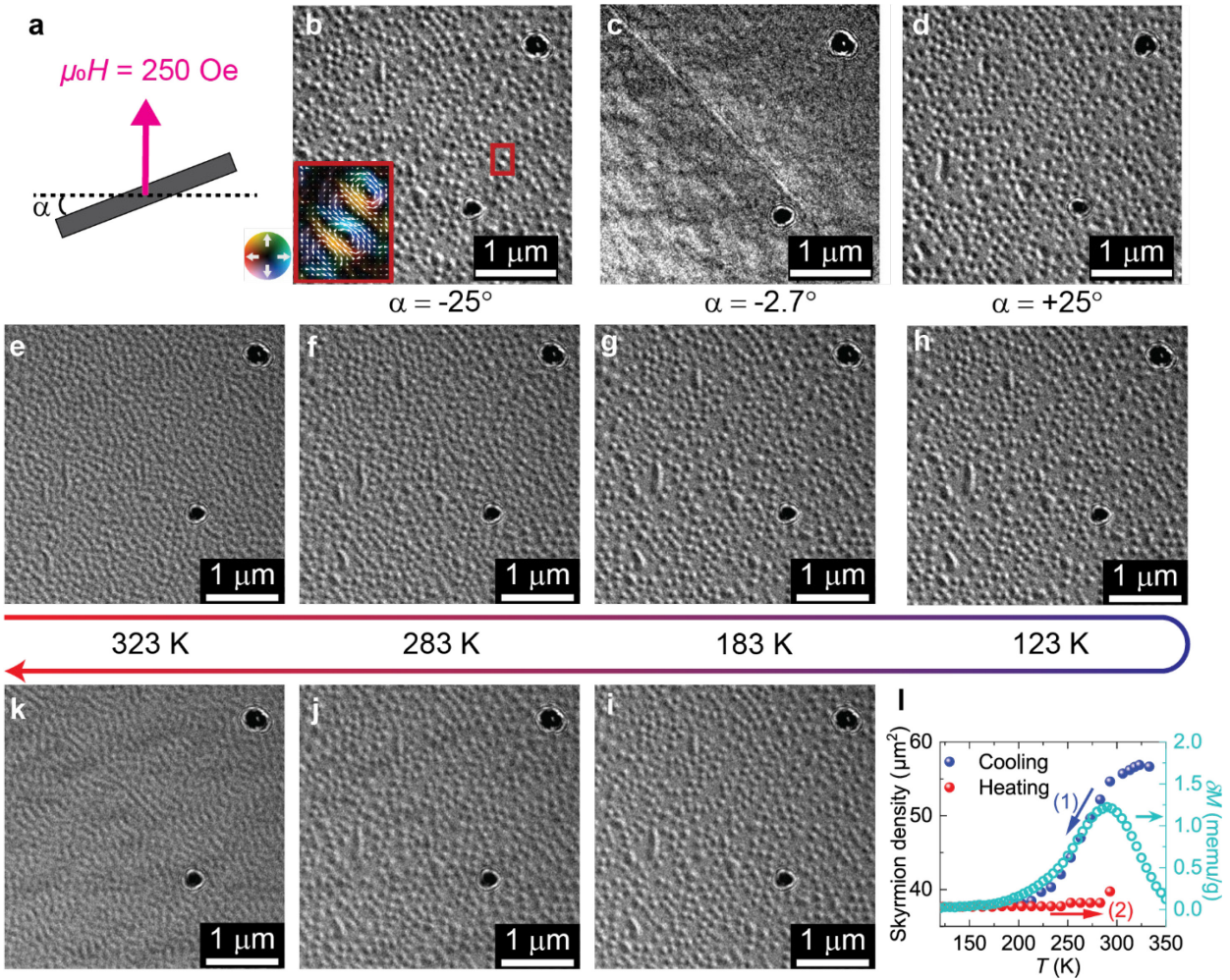

**Figure 5. Lorentz TEM images and magnetization as a function of the temperature. a,** Sketch depicting the experimental configuration used to collect the Lorentz TEM images of Néel-type domains. Sample tilting angle with respect to the horizontal direction is defined by the angle $\alpha$, and the external magnetic field is aligned with the out-of-plane orientation of the sample. Magnetic field is fixed at $\mu_0 H = 250$ Oe during the field cooling process upon which the images were collected. **b, c, and d,** Lorentz TEM images collected at $\alpha = -25°$, $-2.7°$, and $+25°$ at $T = 120$ K, respectively, after field cooling the sample from $T = 373$ K. Magnetic bubbles are only visible at large $\alpha$, indicating that these are Néel-type skyrmions. Inset in **b**: Magnetic induction map of the observed skyrmions showing a winding spin texture. **e** to **k,** Temperature-dependent LTEM images collected at α = −21° when cooling down to $T$ = 123 K from $T$ = 373 K (**e** to **h**) and subsequently warming up from $T$ = 123 K to $T$ = 323 K (**h** to **k**). **l,** Comparison between the temperature-dependent skyrmion density and the size of the thermal hysteresis loop observed in the magnetization, $\delta M$ (cyan open markers). $\delta M$ is obtained by subtracting the magnetization measured on field warming from the one collected upon field cooling.

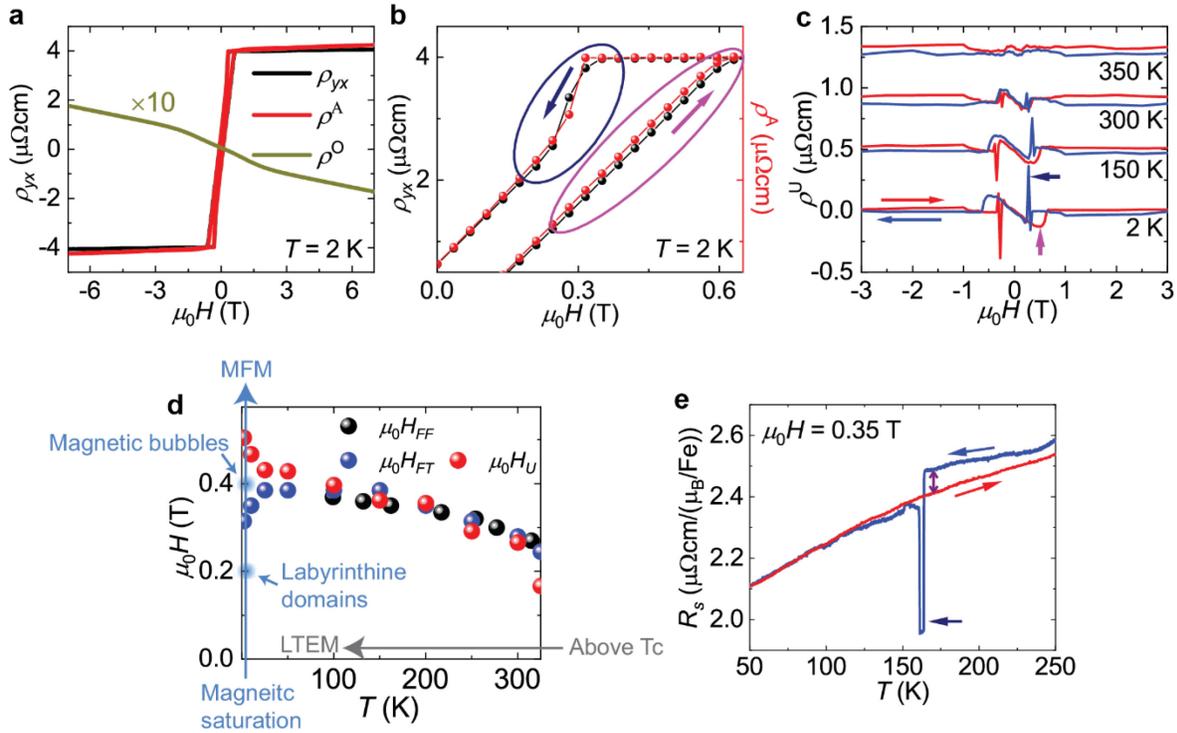

**Figure 6. Analysis of the unconventional component on the Hall effect of $Fe_3GaTe_2$. a,** Anomalous Hall resistivity, $\rho^A$, ordinary Hall resistivity, $\rho^O$, and raw Hall resistivity, $\rho_{yx}$, at $T = 2$ K. Notice that the ordinary Hall-effect $\rho^O$ is non-linear, as expected for a two-carrier system. This is consistent with the band structure calculations shown, for instance, in Ref. [84], indicating the existence of electron and hole Fermi surfaces in $Fe_3GaTe_2$. **b,** Comparison between $\rho_{yx}$ (black markers) and $\rho^A$ (red markers) at $T = 2$ K. Navy and magenta ellipses highlight the field regions where both quantities deviate. Arrows indicate the direction of the field sweep. **c,** Unconventional Hall resistivity $\rho^U$ obtained by subtracting $\rho^A$ and $\rho^O$ from $\rho_{yx}$. Blue (red) lines depict $\rho^U$ measured under decreasing (increasing) magnetic fields. Navy (magenta) arrow indicates the peak in $\rho^U$ in the region marked by the navy (magenta) circle in panel **b,** produced by the transition from the FiM to the spin-polarized state. While the ordinary Hall-effect $\rho^O$ displays nearly linear behavior

for $-1.5 \leq \mu_0 H \leq 1.5$ T, the unconventional Hall $\rho^U$ at $T = 2$ K peaks at $\mu_0 H = 0.5$ T displaying a value that is ~ 3 times larger than that of $\rho^O$. **d,** Critical fields defining the boundary between FM and FiM ($\mu_0 H_{FF}$ and $\mu_0 H_{FT}$) states, as well as the peak position ($\mu_0 H_U$) observed in $\rho^U$ as functions of $T$. Large light blue arrow indicates the path followed for the MFM measurements shown in Figure 4(d-g). Labyrinthine domains and magnetic bubbles are observed under $\mu_0 H = -0.2$ and $-0.4$ T, respectively, and are indicated by light blue circles. Gray arrow indicates the cooling path followed for the LTEM measurements shown in Figure 5(e-h). **e,** $Rs$ as a function of the temperature upon cooling (blue line) and warming (red line) conditions. Spike-like signals likely result from a slight difference in the actual magnetic field values between magnetization and transport measurements as observed in the unconventional Hall effect shown in panel (c). There is a clear decrease in $R_s$ of approximately 3.3% at the FM to FiM transition, as indicated by the purple arrow.

**Methods**

**Sample synthesis**

Fe$_3$GaTe$_2$ single crystals were grown through the chemical vapor transport method. A mix of Fe:Ga:Te in a 6:1:2 ratio was pre-reacted at 1000°C in an evacuated quartz tube for 12 hours. The high Fe content was intended to prevent Fe deficiency. The resulting product was finely ground in an agate mortar. The pre-reacted powder was sealed in an evacuated quartz tube with a small amount of iodine (5 mg/cm$^3$). The tube was placed in a two-zone horizontal tubular furnace with a 750-720°C temperature gradient for two weeks. Subsequently, the tube was naturally cooled down to room temperature. The crystals were washed with acetone and isopropanol. Crystals from a single batch are used in this study.

**Structural and compositional characterization**

Single-crystal diffraction data were obtained at room temperature from a fragment of Fe$_3$GaTe$_2$ (approximately 0.07 x 0.05 x 0.01 mm). Data was collected using a Bruker D8 Quest Kappa single-crystal X-ray diffractometer equipped with an Incoatec I$\mu$s microfocus source (Mo K$_\alpha$ radiation, $\lambda$ = 0.71073 Å) and a PHOTON III CPAD area detector. The raw frames were integrated with Bruker SAINT, and the intensities were corrected for absorption with a multi-scan method in SADABS 2016/2.[85] The intrinsic phasing method in SHELXT was used to generate preliminary crystallographic models,[86] which were finalized with least-squares refinements in SHELXL2019.[87] The crystallographic data and refinement parameters for the model are given in Table S1. Table S2 lists the atomic coordinates, displacement parameters, and refined occupancies. Figure S1 displays the precession images obtained from the refinement of the crystal structure.

TEM lamella was prepared through focused ion beam (FIB) techniques on a ThermoFisher Helios G4 scanning electron microscope (SEM). HAADF-STEM imaging was collected on a probe-aberration-corrected JEOL JEM-ARM200cF with a cold-field emission gun at 200 kV. The STEM resolution of the microscope is 0.78 Å. The HAADF-STEM images were collected with the JEOL HAADF detector using the following experimental conditions: probe size 7c, CL aperture 30μm, scan speed 2 μs/pixel, and camera length 12 cm, which corresponds to a probe semi-convergence angle of 21 mrad. There were about 10 sequential fast-acquired images and they were aligned using a DigitalMicrograph™ plugin.[88] The final images are a sum of aligned 10 images that are free of sample drifting artifacts.

Energy dispersive X-ray spectroscopy was conducted utilizing the Helios UC G4 DualBeam system (Thermo Fisher Scientific) with the Oxford Xmax$^{80}$ EDS (Oxford Instrument) to investigate the chemical composition (see Figure S2). Electron beams with 15 kV and 1.6 nA were used.

**HAADS-STEM image analysis**

Prior to the analysis, the HAADS-STEM signals inferior by 15% to the maximum signal were suppressed to improve the signal to noise ratio. For the Gaussian fittings, a simulated TEM image was generated by initializing position and amplitude parameters. Parameter initialization was conducted based on the lattice information obtained through XRD and a coarse intensity investigation of the TEM image. The difference between the simulated TEM image and the experimental one, was calculated as the reduced-mean-square error with the converged fitting parameters chosen to minimize it (see Figure S3). The code developed for this analysis can be found at https://github.com/Ondue2/TEM_analyzer.

**Electrical and Magnetic properties measurement**

Electrical transport was measured in a commercial Physical Property Measurement System (Quantum Design) under magnetic fields up to $\mu_0 H$ = 9 T and a temperature range of 1.8 ≤ T ≤ 400 K. The Hall resistance was anti-symmetrized while the resistivity was averaged with respect to both field orientations. To deposit the electrical contacts, Ar etching followed by Au deposition was conducted on masked single crystals. Au wires were attached to the gold-deposited areas with Ag paint. Magnetic properties were measured through a commercial Magnetic Properties Measurement System XL Superconducting Quantum Interference Device (Quantum Design) under magnetic fields $-7 \leq \mu_0 H \leq 7$ T and a temperature range of $1.8 \leq T \leq 400$ K.

**Neutron Diffraction measurements**

Elastic neutron scattering measurements on a ≈ 5 mg single crystal sample of $Fe_3GaTe_2$ were performed in zero applied magnetic field at the University of Missouri Research Reactor (MURR) using the triple-axis spectrometer TRIAX. An incident beam energy of 14.7 meV was selected with a pyrolytic graphite (PG) monochromator, and a PG analyzer after the sample was tuned to the same energy to reduce the background. PG filters were positioned in the incident and scattered beams to remove higher wavelength harmonics. The beam divergence was limited with 60′-60′-80′-80′ collimators positioned, respectively, before the monochromator, between the monochromator and sample, between the sample and analyzer, and between the analyzer and detector. The sample was sealed in a $^4$He environment within an aluminum can and subsequently mounted on the cold tip of an Advanced Research Systems closed-cycle refrigerator. The crystal was aligned in the (*h* 0 *l*) scattering plane that provided access to key structural and magnetic reflections.

Neutron scattering measurements under magnetic fields were performed at the High Flux Isotope Reactor (HFIR) at the Oak Ridge National Laboratory (ORNL). $\lambda = 1.486$ Å wavelength was used with a vertically focused and hot-pressed Ge monochromator, with the scattered neutron beam collected using the Wide-Angle Neutron Diffractometer (WAND$^2$) (curved, $^3$He 2D-PSD with

~$2e^6$ pixels, within the scattering angles of 0° < $2\theta$ < 156°). The sample environment was controlled in a liquid helium cryostat ($T$ = 1.8 – 300 K) and a cryomagnet ($\mu_0H$ = –6 – 6 T). The sample was attached to an Al sample holder having the *ab*-plane parallel to the incident neutron beam, with the magnetic field applied along the *c*-axis. The Al sample holder was covered with $Gd_2O_3$ foil to minimize the signal from the sample holder. The sample alignment was fine-tuned by tracking the (110) peak intensity. For the temperature-dependent experiment, the angle of incident beam was fixed to the maximize the Bragg reflection along the (110) peak at $T$ = 290 K. For the measurements at $T$ = 100 K under $\mu_0H$ = 0.35 and 1 T, a full scan was collected by rotating the sample 180° around its *c*-axis.

**LTEM and MFM measurement**

Single-crystalline $Fe_3GaTe_2$ was mechanically exfoliated directly onto a homemade polydimethylsiloxane stamp inside an argon-filled glovebox. Prior to its utilization, the stamp was rinsed in acetone and isopropyl alcohol to clean its surface. After appropriate crystal thicknesses and dimensions were identified via optical contrast, the selected crystal(s) was transferred onto one window of a silicon-nitride-based TEM grid. A 2-3 nm layer of Pt was deposited on the transferred flake using DC sputtering. Cryogenic LTEM measurements was performed in a JEOL transmission electron microscope in Lorentz mode using a Gatan HC3500 heating-cooling TEM holder. A perpendicular magnetic field aligned parallel to the electron beam was generated by applying a small amount of current to the main objective lens. All LTEM images were collected through the underfocus condition with a defocus length of approximately 2.5 mm. The magnetic induction maps were reconstructed from the out-of-focus LTEM image based on transport-of-intensity equation.[89]

Magnetic force microscopy (MFM) measurements were performed using a home-built $^4$He set-up.[90] All the measurements used a commercial MFM tip (PPP-MFMR, NANOSENSORSTM). The magnetic contrast in MFM images is expressed by the frequency shift $\Delta f$ of the MFM cantilever, which is directly linked to the force gradient $\frac{\partial F}{\partial z}$ with the relation $\Delta F = -\frac{f_0}{2k}\frac{\partial F}{\partial z}$, where $k$ (2.5 N/m) is the force constant and $f_0$ (71.234 kHz) is the resonance frequency of the MFM cantilever. The coercivity of the MFM tip along the out-of-plane direction at $T = 4.25$ K is approximately 800 G.

**Micromagnetic simulation**

Micromagnetic simulation was conducted via MUMAX.[91] Exchange stiffness of $7.5\times10^{-12}$ J/m, first order uniaxial anisotropy of $3\times10^5$ J/m$^3$, interfacial DM interaction strength of $1.07\times10^{-3}$ J/m$^2$, Landau-Lifshitz damping constant of 0.2, and external magnetic field of 800 Oe were used for the MUMAX parameters. The value of exchange stiffness was taken from Ref. [41], and the values of interfacial DM interaction and Lifshitz damping constant were taken from Ref. [39]. The value of uniaxial anisotropy was reduced from the experimental value of $7.9\times10^5$ J/m$^3$ at 300 K to $3.0\times10^5$ J/m$^3$ to nucleate skyrmions. The experimental value of $2.95\times10^5$ A/m at 300 K was used for the saturation magnetization parameter, which generated Néel-type skyrmions. A significantly enhanced saturation magnetization parameter of $4.8\times10^5$ J/m$^3$ was used to simulate Bloch-dominated hybrid skyrmions.

**Supporting information:** Supporting Information is available free of charge at https://pubs.acs.org /doi/. Single-crystal XRD precession images, EDS data, HAADF-STEM images, probability distributions of atomic displacements with respect to the mirror plane between Te atoms as obtained from fittings of the TEM image, derivatives of magnetic susceptibility with respect to temperature, neutron scattering, temperature-dependent magnetic anisotropy analysis, temperature-dependent anomalous Hall conductivity and resistivity data upon field-cooling and warming, Mean-field model parameters for the FM to FiM transition, field-dependent transverse and longitudinal resistivity, topological Hall signal peak finding procedure, LTEM images under different magnetic fields collected upon cooling, micromagnetic simulations, reconstructed LTEM images, difference in $S_H$ between FM and FiM phases, the absence of the *ab*-plane Hall effect, crystallographic tables, and movies exposing the temperature dependence of the skyrmion density.

**ACKNOWLEDGEMENTS**

L.B. acknowledges support from the US DoE, BES program through award DE-SC0002613 US (synthesis and measurements), US-NSF-DMR 2219003 (heterostructure fabrication) and the Office Naval Research DURIP Grant 11997003 (stacking under inert conditions). J.Y.C. acknowledges DOE DE-SC0022854 and the Welch Foundation through AA-2056-20240404. The National High Magnetic Field Laboratory acknowledges support from the US-NSF Cooperative agreement Grant DMR-2128556, and the state of Florida. Work at Argonne (Y.L., C.P.) was funded by the US Department of Energy, Office of Science, Office of Basic Energy Sciences, Materials Science and Engineering Division. Use of the Center for Nanoscale Materials, an Office of Science user facility, was supported by the U.S. Department of Energy, Office of Science, Office of Basic Energy Sciences, under Contract No. DE-AC02-06CH11357. Jeehoon Kim acknowledges the support from the National Research Foundation of Korea (NRF) grant funded by the Korean government (MSTI) (No. RS-2024-00410027). Jaeyong Kim acknowledges the support from the National Research Foundation of Korea (NRF) grant funded by the Korean government (MSTI) (No. 2022H1D3A3A01077468). E.J.G.S. acknowledges computational resources through CIRRUS Tier-2 HPC Service (ec131 Cirrus Project) at EPCC (http://www.cirrus.ac.uk), which is funded by the University of Edinburgh and EPSRC (EP/P020267/1); and ARCHER2 UK National Supercomputing Service via the UKCP consortium (Project e89) funded by EPSRC grant ref EP/X035891/1. E.J.G.S. acknowledges the EPSRC Open Fellowship (EP/T021578/1), the Donostia International Physics Center for funding support. PC acknowledges funding support from the China Scholarship Council grant (202208060246). This research used resources at the High Flux Isotope Reactor, a Department of Energy (DOE) Office of Science User Facility operated by Oak Ridge National Laboratory (ORNL). S.-E. L. was partially supported by the National Research Foundation of Korea (NRF) grant (Nos. RS-2024-00412446).

**Disclaimer**

We identify certain commercial equipment, instruments, and materials in this article to specify adequately the experimental procedures. In no case does such identification imply recommendation or endorsement by the National Institute of Standards and Technology nor does it imply that the materials or equipment identified are necessarily the best available for the purpose.

**Conflicts of Interest**

The authors declare no conflicts of interest.

**Data Availability Statement**

The data that support the findings reported in this manuscript can be accessed via this DOI: 10.17605/OSF.IO/SW8CA

# Supporting Information for manuscript titled: Local Inversion Symmetry Breaking and Thermodynamic Evidence for Ferrimagnetism in $Fe_3GaTe_2$

Sang-Eon Lee[1,7], Yue Li[3], Yeonkyu Lee[4], W. Kice Brown[5], PeiYu Cai[6], Jinyoung Yun[4], Chanyoung Lee[4], Alex Moon[1,2], Lingrui Mei[1,2], Jaeyong Kim[7], Yan Xin[1], Julie A. Borchers[8], Thomas W. Heitmann[9,10,11], Matthias Frontzek[12], William D. Ratcliff[8,13,14], Gregory T. McCandless[5], Julia Y. Chan[5], Elton J. G. Santos[6,15,16], Jeehoon Kim[4], Charudatta M. Phatak[3], Vadym Kulichenko[1], Luis Balicas[*1,2]

1. National High Magnetic Field Laboratory, Tallahassee, Florida 32310, United States

2. Department of Physics, Florida State University, Tallahassee, Florida 32306, United States

3. Materials Science Division, Argonne National Laboratory, Lemont, Illinois 60439, United States

4. Department of Physics, Pohang University of Science and Technology, Pohang 37673, Republic of Korea

5. Department of Chemistry & Biochemistry, Baylor University, Waco, Texas 76706, United States

6. Institute for Condensed Matter and Complex Systems, School of Physics and Astronomy, The University of Edinburgh, Edinburgh EH9 3FD, U.K.

7. Department of Physics, Hanyang University, Seoul 04763, Republic of Korea

8. NIST Center for Neutron Research National Institute of Standards and Technology Gaithersburg, MD 20899, USA.

9. University of Missouri Research Reactor, University of Missouri, Columbia, Missouri 65211, USA

10. Department of Physics and Astronomy, University of Missouri, Columbia, Missouri 65211, USA

11. MU Materials Science and Engineering Institute, University of Missouri, Columbia, Missouri 65211, USA

12. Neutron Scattering Division, Oak Ridge National Laboratory, Oak Ridge, Tennessee 37831, USA

13. Department of Physics, University of Maryland, College Park, Maryland 20742, United States

14. Department of Materials Science and Engineering, University of Maryland, College Park, Maryland 20742, United States

15. Higgs Centre for Theoretical Physics, The University of Edinburgh, Edinburgh EH9 3FD, U.K.

16. Donostia International Physics Center (DIPC), 20018 Donostia-San Sebastián, Basque Country, Spain

*balicas@magnet.fsu.edu

**Supporting information includes:**

Figure S1. *hk*0, *h*0*l*, and 0*kl* precession images for $Fe_3GaTe_2$.

Figure S2. Energy dispersion spectroscopy (EDS) from $Fe_3GaTe_2$.

Figure S3. Gaussian fits to the high-angle annular dark-field scanning transmission electron microscopy (HAADF-STEM) images.

Figure S4. Probability distributions of atomic displacements with respect to the mirror plane between Te atoms as obtained from fittings of the TEM image.

Figure S5. Intercalated Fe3 atoms observed via HAADF-STEM imaging.

Figure S6. Derivatives of magnetic susceptibility with respect to temperature.

Figure S7. Elastic neutron diffraction measurements of structural and magnetic reflections as a function of temperature for a $Fe_3GaTe_2$ crystal.

Figure S8. Temperature-dependent magnetic anisotropy analysis for $Fe_3GaTe_2$.

Figure S9. Hall conductivity and resistance for $Fe_3GaTe_2$ as functions of the temperature under fixed magnetic fields.

Figure S10. Size difference between demagnetization field and external magnetic field at the ferromagnetic to ferrimagnetic transition.

Figure S11. Resistivity of $Fe_3GaTe_2$ as a function of the magnetic field.

Interpretation of the MFM images

Figure S12. Dependence of the MFM tip frequency shift on the magnetic susceptibility.

Figure S13. LTEM images collected under different magnetic fields after the cooling procedure.

Figure S14. Micromagnetic simulations and reconstructed LTEM images for Néel and Bloch-dominated hybrid skyrmions.

Figure S15. Unconventional Hall effect and phase boundary in lower-quality samples.

Figure S16. Procedure used to extract the position of the maxima in the topological Hall signal.

Figure S17. $S_H$ as a function of the temperature upon cooling (blue line) and warming (red line) under $\mu_0 H = 0.35$ T.

Figure S18. Hall resistivity for a $Fe_3GaTe_2$ single crystal at $T = 2$ K (left) and $T = 300$ K (right) as a function of the magnetic field and for several field orientations.

Table S1.  Crystallographic Data, Data Collection, and Refinement Parameters (Single Crystal X-ray Diffraction).

Table S2. Fractional atomic coordinates and displacement parameters (Single Crystal X-ray Diffraction).

Table S3. Comparison between lattice parameters and Fe2/Fe3 site occupancies for different $Fe_3GaTe_2$ samples.

Table S4. Stoichiometry of a crystal according to EDS.

Movie S1. LTEM images collected on cooling temperature.

Movie S2. LTEM images collected on warming temperature.

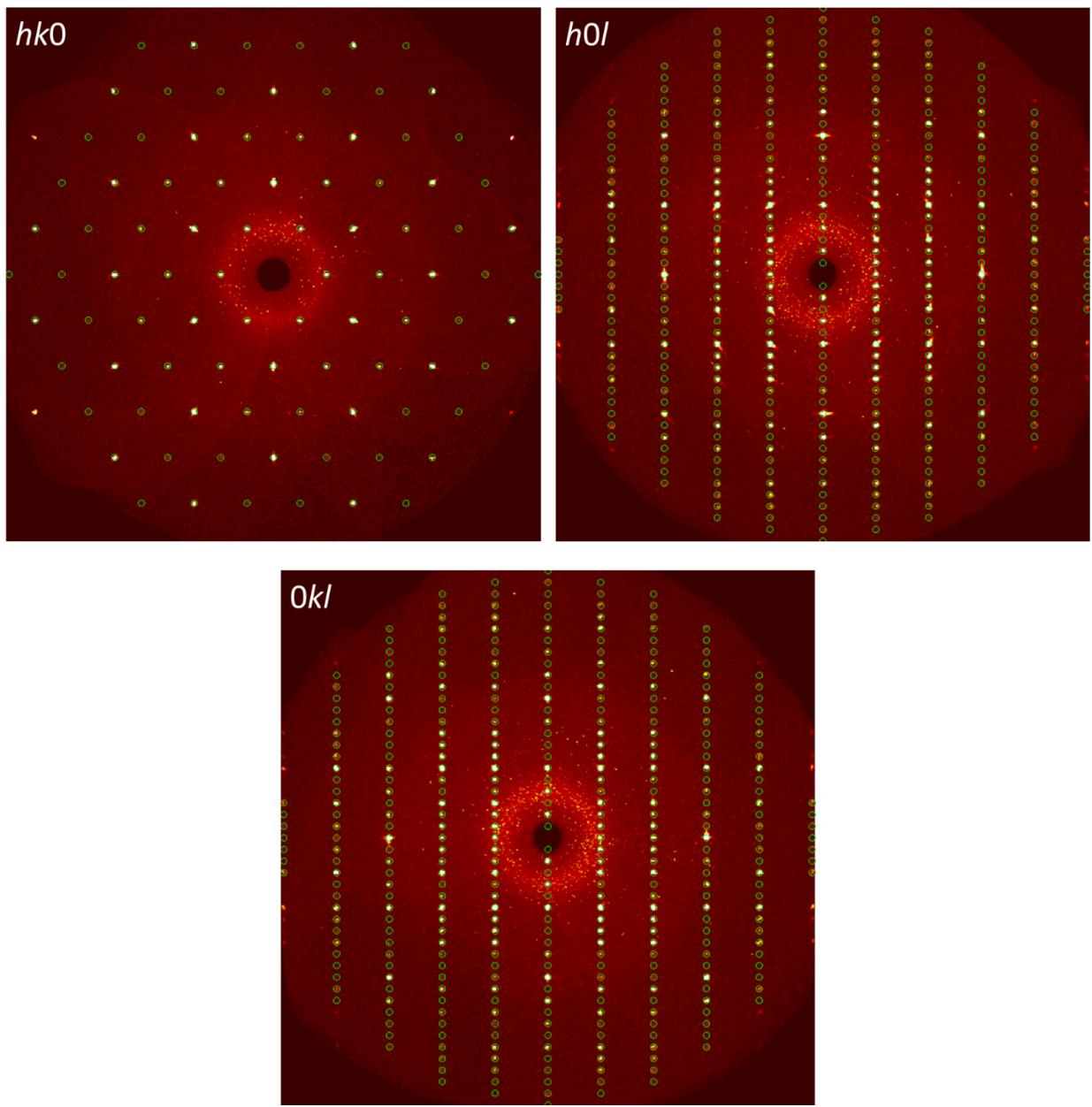

**Figure S1. *hk*0, *h*0*l*, and 0*kl* precession images for Fe₃GaTe₂.** Reflections are indexed with a hexagonal unit cell ($a$ = 4.0793(3) Å, $c$ = 16.107(3) Å, $V$ = 232.12(5) Å$^3$) shown with the green indexing circles.

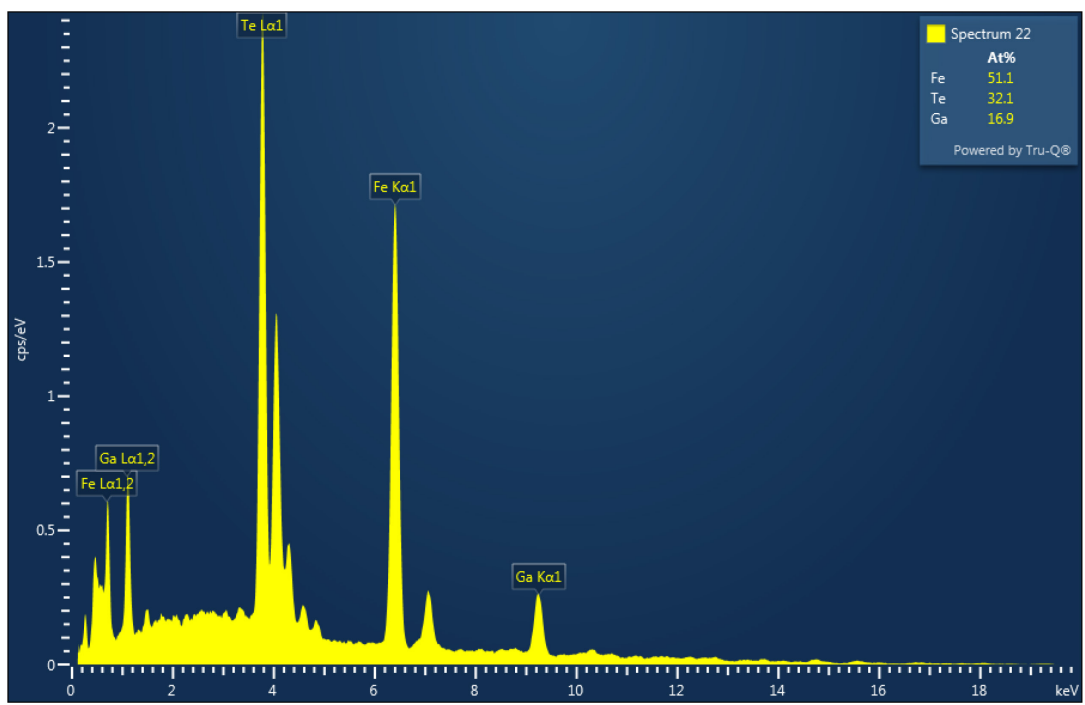

**Figure S2. Energy dispersion spectroscopy (EDS) from Fe$_3$GaTe$_2$.** The average composition from five EDS spectra yields the ratio Fe:Ga:Te = 3.04(5):0.96(1):2.0(3) with the Te composition fixed to a value of two. The compositions from each data point are shown below in Table S4.

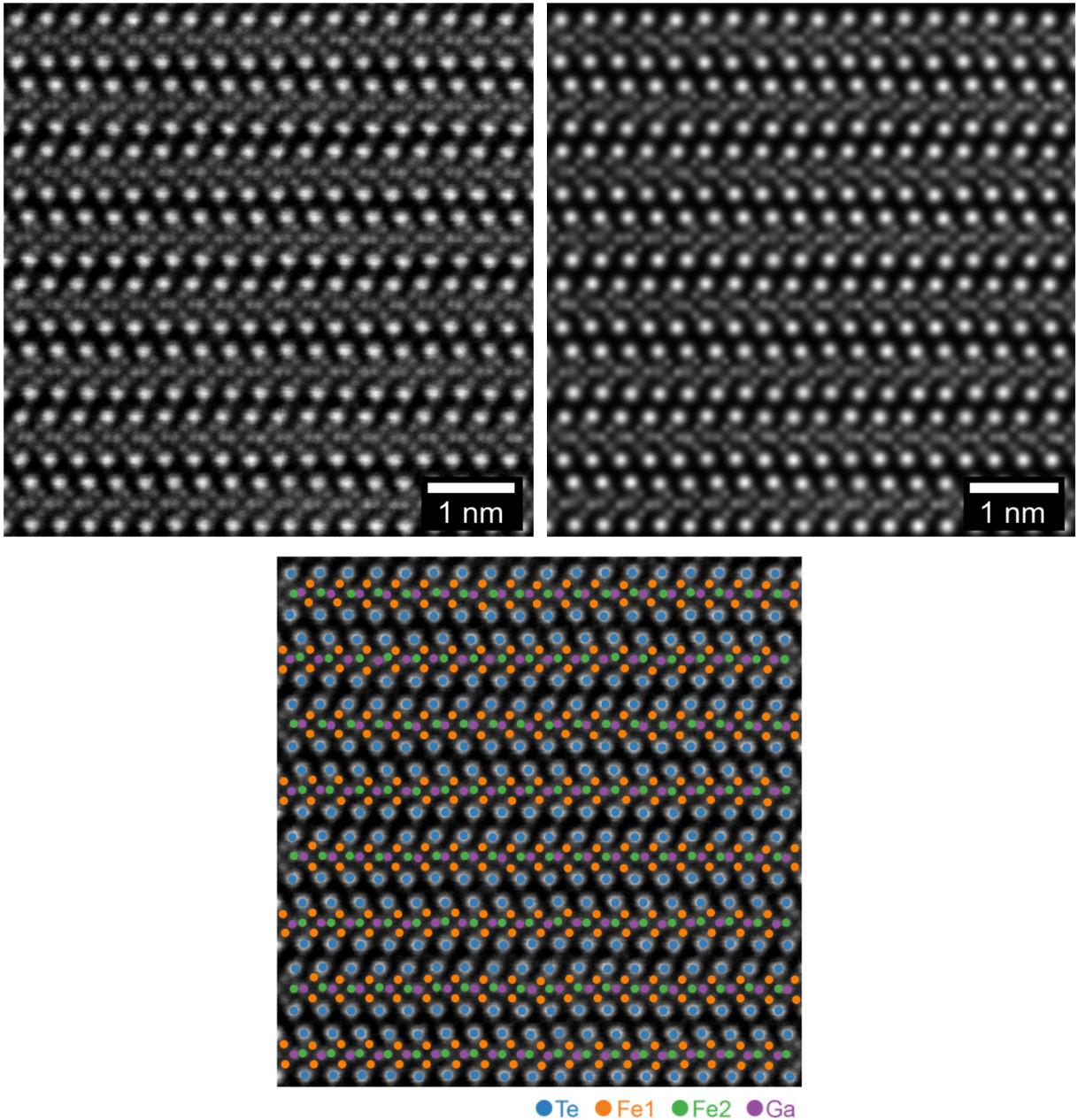

**Figure S3. Gaussian fits to the High-Angle Annular Dark-Field Scanning Transmission Electron Microscopy (HAADF-STEM) images.** Experimental and fitted HAADF-STEM images using a Gaussian function are shown at the left and right of the upper panel, respectively. Experimental image with superimposed atomic coordinate markers, whose coordinates were obtained from the Gaussian fits, is shown in the lower panel. Blue, orange, green, and purple circles represent Te, Fe1, Fe2, and Ga atoms, respectively.

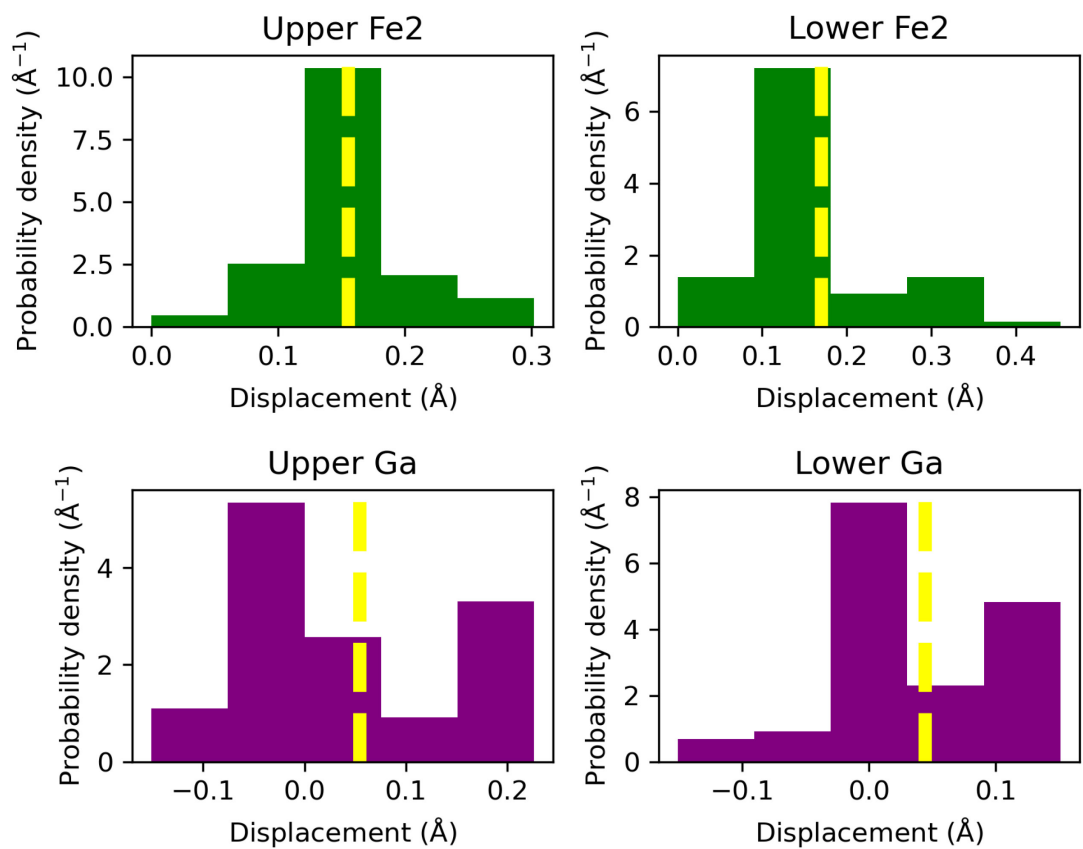

**Figure S4. Probability distributions of atomic displacements with respect to the mirror plane between Te atoms as obtained from fittings of the TEM image.** Dashed yellow lines indicate the average displacements.

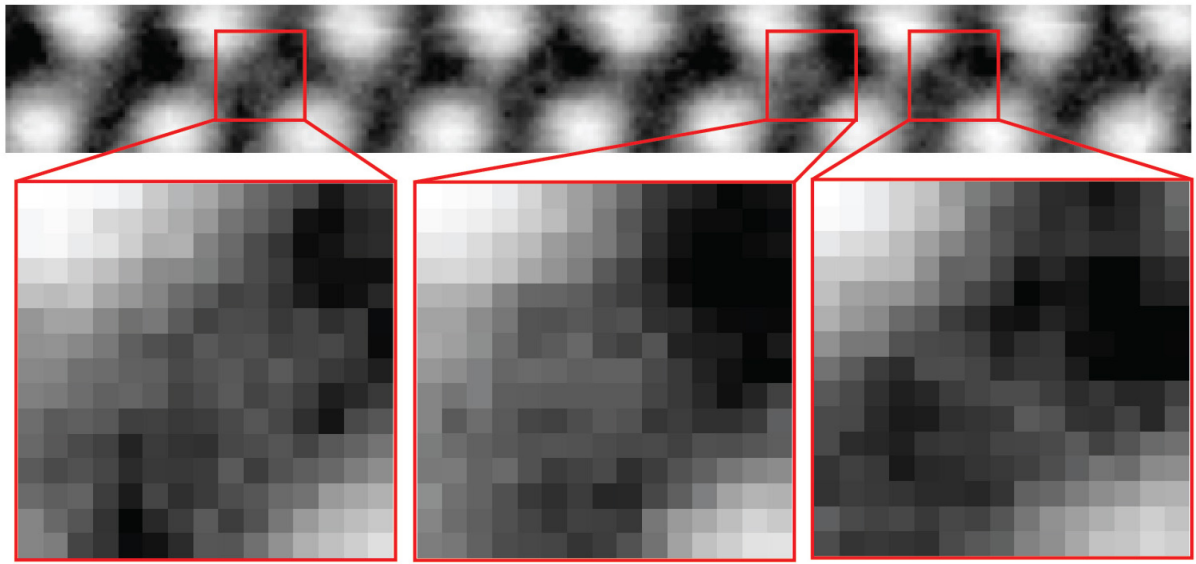

**Figure S5. Intercalated Fe3 atoms observed via HAADF-STEM imaging.** Magnified view of the van der Waals gap region of $Fe_3GaTe_2$ via HAADF-STEM imaging. Weak signals corresponding to intercalated Fe atoms, i.e, Fe3, which can be identified in the regions contained within the red boxes. In this image, signals inferior by 5% to the maximum signal were suppressed to improve the signal to noise ratio. To enhance the contrast, a logarithmic scale was applied to the signal.

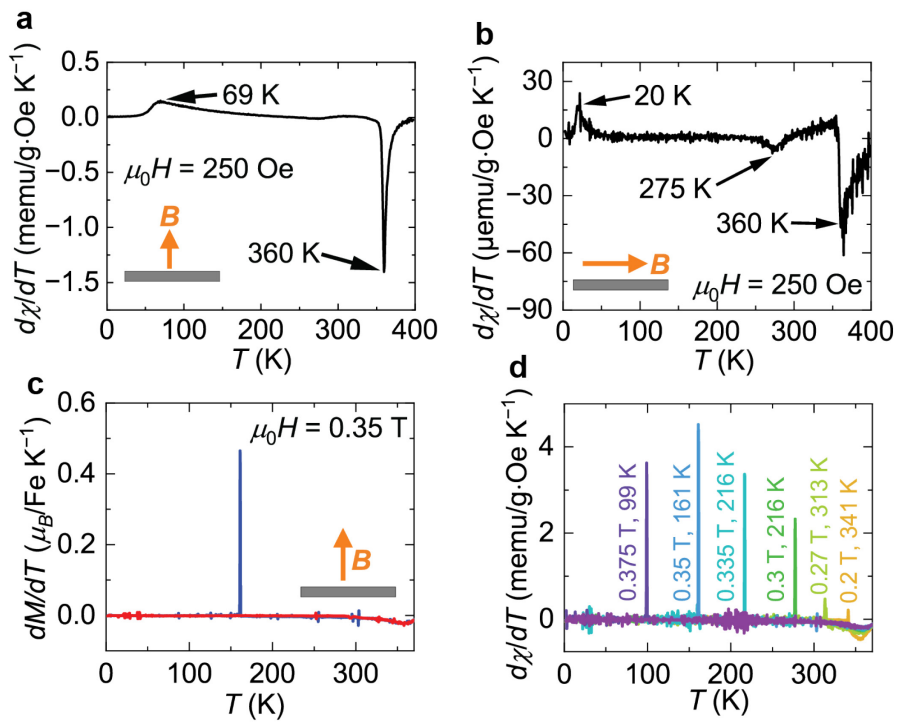

**Figure S6. Derivatives of magnetic susceptibility with respect to temperature. a** and **b**, Derivatives of ZFC susceptibility traces as functions of the temperature collected under $\mu_0 H = 250$ Oe for different magnetic field orientations. **c**, Derivatives of the magnetization as a function of $T$ measured under $\mu_0 H = 0.35$ T upon cooling (the blue line) and warming (the red line). **d**, Derivatives of the temperature-dependent magnetic susceptibilities measured under various magnetic fields.

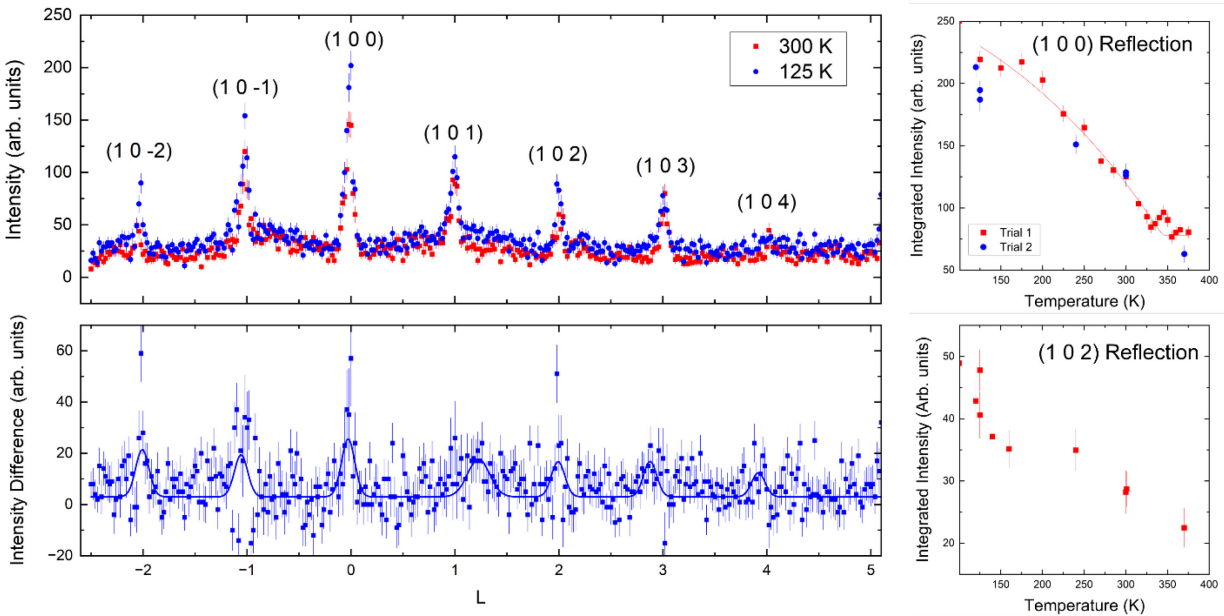

**Figure S7. Elastic neutron diffraction measurements of structural and magnetic reflections as a function of temperature for a Fe$_3$GaTe$_2$ crystal.** Neutron diffraction measurements were performed in zero field as a function of temperature in the (*h 0 l*) scattering plane. The upper left graph shows scans at 300 K (near the magnetic ordering temperature) and at 125 K along the (1 0 *l*) direction. Note that higher order reflections up to (1 0 8) were detectable. The peaks in the 300 K data are primarily structural, and the lattice parameters extracted from the peak positions [and from the (0 0 2) peak position] are consistent with those obtained from x-ray analysis (SCXRD), confirming *P*6$_3$/*mmc* space group symmetry. Additional magnetic scattering develops at the (1 0 *l*) peak positions upon cooling to 125 K. The magnetic contributions to the scattering at each peak position can be isolated by subtracting the 300 K data from the 125 K data, as shown in the bottom left figure. In contrast, (0 0 *l*) scans through the (0 0 2) reflection show no significant temperature dependence within error. As a reminder, neutron selection rules indicate that only the magnetization perpendicular to the scattering vector contributes to the scattering process. These neutron scattering data thus confirm that magnetic order is collinear along the c-axis with no detectable canting. The scattering is consistent with either a ferromagnetic or ferrimagnetic alignment of the moments, and current fit attempts cannot distinguish between the two possibilities. Measurements of the magnetic order parameter were obtained by performing rocking curves at the (1 0 0) and (1 0 2) peak positions upon heating, shown in the upper right and lower right graphs

respectively. Both curves approximately track the temperature dependence of the magnetization upon heating (Fig 2a), though the neutron intensity is proportional to the square of the magnetization. A fit of a Brillouin function to the (1 0 0) data reveals an ordering temperature of 343 ± 2 K, which is consistent with the Curie temperature extracted from bulk magnetization results.

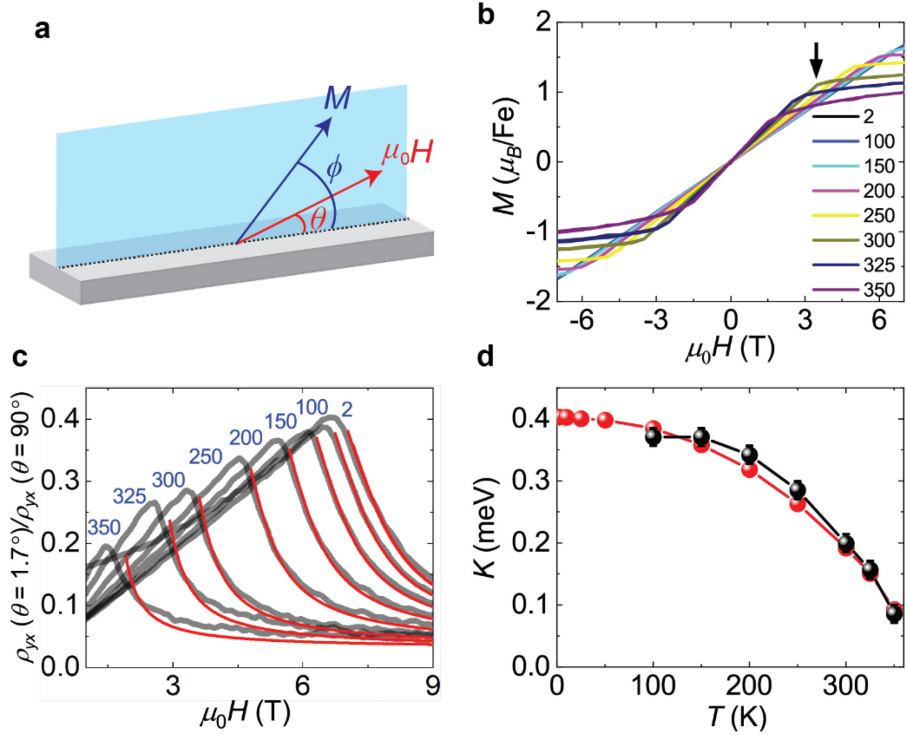

**Figure S8. Temperature-dependent magnetic anisotropy analysis for Fe$_3$GaTe$_2$.** We described the magnetic field orientation-dependent magnetic energy within a simple model of $E = -\mu_0\mu_{eff}H\cos(\phi - \theta) + K\cos^2\phi$, where $\mu_{eff}$ is the effective magnetic moment, $K$ is the magnetic anisotropy energy, and $\phi$ ($\theta$) is the angle between the *ab*-plane and the magnetization (magnetic field). We used $\mu_{eff} = 1.97$ $\mu_B$, as extracted from the magnetization under $\mu_0 H = 7$ T at $T = 2$ K for magnetic fields applied along the *c*-axis. Configuration is shown in **a**. **b,** In-plane magnetization of Fe$_3$GaTe$_2$. The magnetization displays kinks where its slope changes. Black arrow represents the kink at $T = 300$ K. In the model, the magnetic anisotropy energy can be estimated as $K = \mu_0\mu_{eff}H_k/2$, where $\mu_0 H_k$ is the magnetic field where the kink appears. **c,** Normalized Hall resistivities evaluated by dividing the Hall resistivities at $\theta = 1.7°$ by the Hall resistivities at $\theta = 90°$ (black solid lines). The measurement was conducted with a decreasing magnetic field from $\mu_0 H = 9$ T. Blue numbers indicate temperatures in K. This normalized Hall resistivity approximately represents $\sin\phi$ in the single-domain regime since the Hall resistivity, $\rho_{yx}$, only depends on the *c*-axis component of the magnetization, $M\sin\phi$. From the model, we evaluated $\phi$ by minimizing the magnetic energy and fitted the data (red solid lines). **d,** Evaluated magnetic

anisotropic energy per Fe atom, $K$, as a function of the temperature. Black (red) symbols display the values of $K$ estimated from the in-plane magnetization (normalized Hall resistivities). $K$ values from the in-plane magnetization are limited to high temperatures since we could not observe the kink in the low-temperature region under magnetic fields limited to $\mu_0 H = 7$ T. We extrac $K$ ~ 0.40 meV at $T = 2$ K and ~ 0.20 meV at $T = 300$ K.

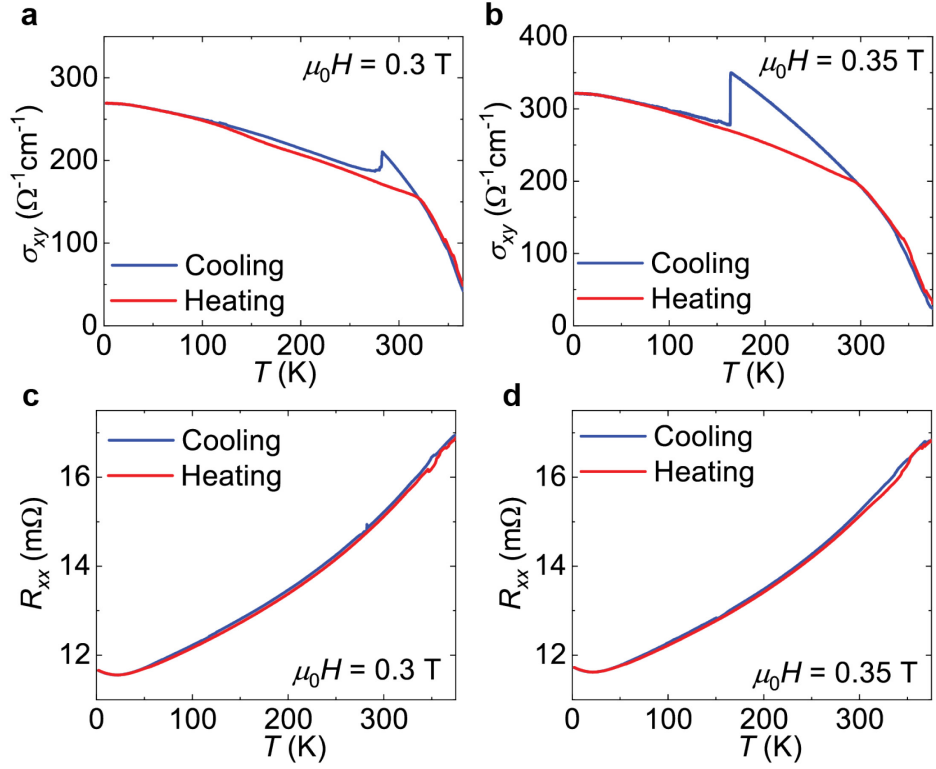

**Figure S9. Hall conductivity (a, b) and resistance (c, d) for $Fe_3GaTe_2$ as functions of the temperature under fixed magnetic fields.** Strengths of the applied magnetic field were $\mu_0 H =$ 0.3 T (left panels) and $\mu_0 H = 0.35$ T (right panels). A sudden drop is seen in the anomalous Hall conductivity, as well as thermal hysteresis, both resembling those observed in the magnetization. Magnetic transitions are observed only in the anomalous Hall conductivity.

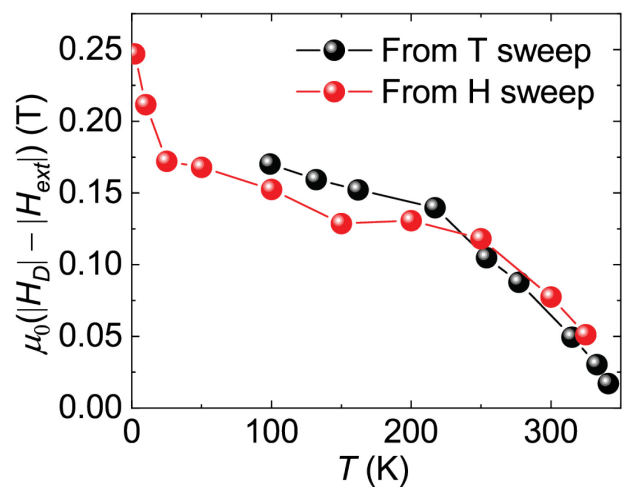

**Figure S10. Difference between demagnetization and external magnetic fields at the ferromagnetic to ferrimagnetic transition.** The demagnetization field is calculated with a demagnetization factor = 1. The size of $H_D$ is always larger than that of $H_{ext}$, suggesting that the demagnetization field contributes to the ferromagnetic to ferrimagnetic transition.

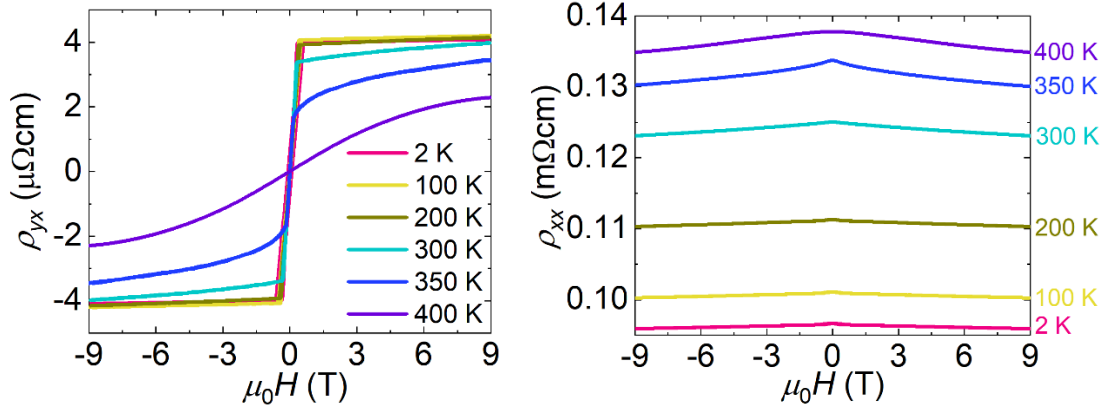

**Figure S11. Resistivity of Fe$_3$GaTe$_2$ as a function of the magnetic field.** Hall resistivity, $\rho_{yx}$, and the longitudinal magnetoresistivity, $\rho_{xx}$, as functions of the magnetic field shown in the left and right panels, respectively. The field dependence of $\rho_{yx}$ follows that of the magnetization. $\rho_{xx}$ decreases slightly upon increasing magnetic fields, due to the magnetic field induced suppression of spin scattering.

**Interpretation of the MFM images**

The MFM is a useful tool for investigating the magnetization of samples, but the relation between the MFM signal and the sample's magnetization is not straightforward. In this note, we summarized the theory of MFM and interpreted the dark blue region observed by us (Figure 4 in the main text). First, it is well known that the frequency shift of the MFM tip can be approximately linearly proportional to the force gradient as $\Delta f \approx -\partial F_z/\partial z$. The potential energy of the magnetic dipole-dipole interaction between the two-point like magnetic moments $\mathbf{m}_1$ and $\mathbf{m}_2$ is given by

$$H = -\frac{\mu_0}{4\pi}\left[\frac{3(\mathbf{m}_1 \cdot \mathbf{r})(\mathbf{m}_2 \cdot \mathbf{r})}{r^5} - \frac{\mathbf{m}_1 \cdot \mathbf{m}_2}{r^3}\right], \quad (1)$$

where $\mu_0$ is the vacuum magnetic permeability, and $\mathbf{r}$ is the vector connecting the relative positions of $\mathbf{m}_1$ and $\mathbf{m}_2$. Here, we assume the tip and sample's magnetization have only $z$-components, the tip magnetization is approximated to a point magnetic moment, and the sample is two-dimensional and infinitely wide for simplicity. In this case, the force that the MFM tip feels from the sample magnetization can be obtained straightforwardly from the magnetic dipole-dipole interaction.

$$F_z = -\frac{\partial H}{\partial z} = \frac{3\mu_0 m_t z}{2}\int_0^\infty \frac{M(\rho)}{r^5}\left[3-\left(\frac{z}{r}\right)^2\right]\rho d\rho, \quad (2)$$

where $m_t$ is the tip magnetic moment, $M$ is the sample magnetization, $z$ is the distance between the tip and sample surface, and $\rho$ is the distance between the $xy$ position of the tip and the specific sample region (see Figure S12a for the configuration). Finally, the frequency shift can be approximated as follows.

$$\Delta f \approx -\frac{\partial F_z}{\partial z} \propto -m_t \int_0^\infty \frac{M(\rho)}{r^5}\left[3-30\left(\frac{z}{r}\right)^2+35\left(\frac{z}{r}\right)^4\right]\rho d\rho. \quad (3)$$

If $M$ is constant with respect to $\rho$, the rightest term becomes zero. Therefore, the MFM is a tool for investigating the spatial variation of the magnetization rather than the magnetization itself, which is useful for domain investigation. If the magnetization of the sample becomes weaker (stronger) as the distance from the tip increases, the frequency will shift negatively (positively).

In the case that the magnetic susceptibility of the sample is large enough to respond to the magnetic field of the tip, positive (negative) magnetic susceptibility leads to a negative (positive) frequency shift. This can be demonstrated with a magnetization that is linearly proportional to the z-component of the magnetic field produced by the tip as $M = \chi \frac{\mu_0 m_t}{4\pi} \frac{1}{r^3}\left[3\left(\frac{z}{r}\right)^2 - 1\right]$, where $\chi$ is the magnetic susceptibility. If we plug this expression into Eq. (3), we obtain:

$$\Delta f \propto -\chi \int_0^\infty \frac{1}{r^8}\left[3\left(\frac{z}{r}\right)^2 - 1\right]\left[3 - 30\left(\frac{z}{r}\right)^2 + 35\left(\frac{z}{r}\right)^4\right]\rho d\rho \propto -\chi. \qquad (4)$$

Note that the integral of the second term should yield a positive constant. Figures S11b and S11c illustrate respectively, the cases for small and large magnetic susceptibilities, as well as the response of the spin configuration with respect to the magnetic field of the tip. For a small magnetic susceptibility, the magnetic field from the tip cannot change the ferromagnetic configuration, thus producing negligible frequency shifts. As for the large magnetic susceptibility case, the magnetic moment configuration is sensitive to the stray field from the tip, producing a large negative frequency shift. Each case corresponds to the blue and dark blue colored regions in the MFM images (Figure S12d).

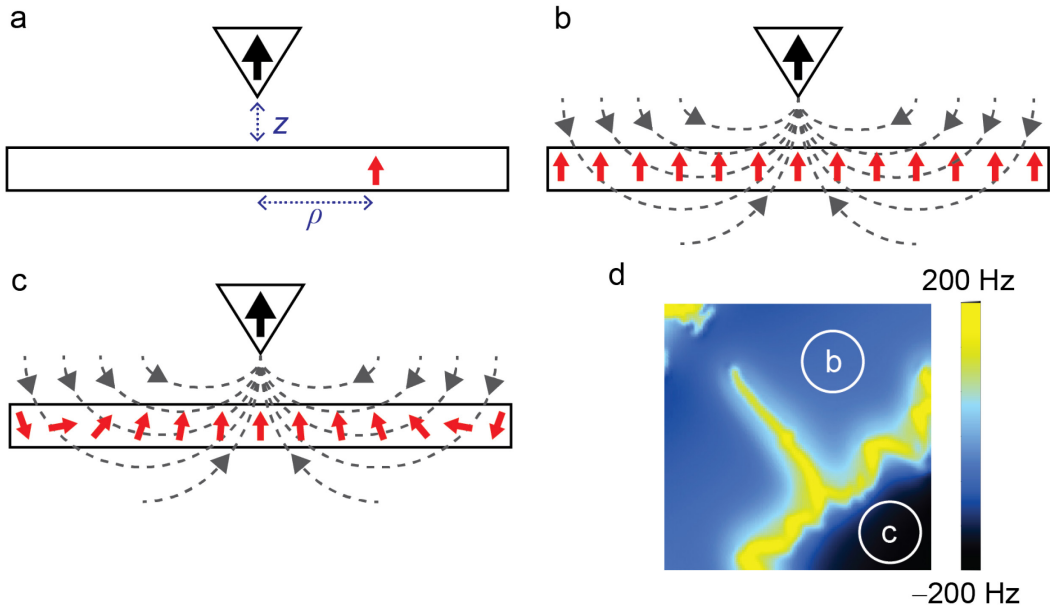

**Figure S12. Dependence of the MFM tip frequency shift on the magnetic susceptibility. a**, MFM tip magnetization, and sample configuration whose parameters are used in the above discussion. **b**, Spin configuration of the small magnetic susceptibility sample when affected by the stray field of the tip. **c**, Spin configuration of the large magnetic susceptibility sample when affected by the stray field of the tip. **d**, MFM image of $Fe_3GaTe_2$ (Figure 4e within the main manuscript). The (dark) blue region corresponds to the large (small) magnetic susceptibility region.

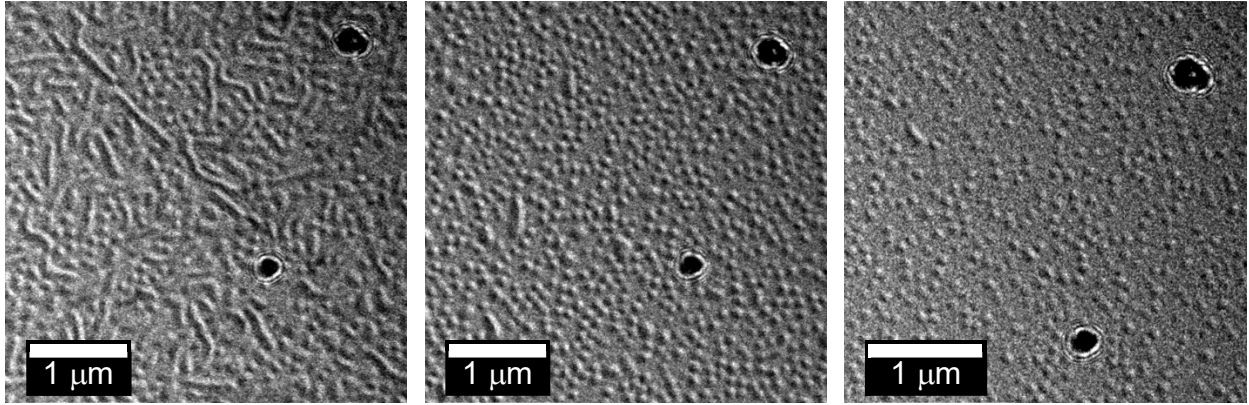

**Figure S13. LTEM images collected under different magnetic fields after cooling.** Left, middle, and right panels correspond to magnetic fields of $\mu_0 H$ = 155, 250, and 350 Oe and tilting angles of $\alpha$ = −21, −21, and −23° respectively, applied to the sample after cooling. This measurement was conducted at $T$ = 223 K. Stripe domains are favored by the smaller magnetic field of $\mu_0 H$ = 155 Oe. A lower skyrmion density is observed under a higher magnetic field of $\mu_0 H$ = 350 Oe (right panel), when compared to the one observed under $\mu_0 H$ = 250 Oe (middle panel) since the system approaches magnetic saturation.

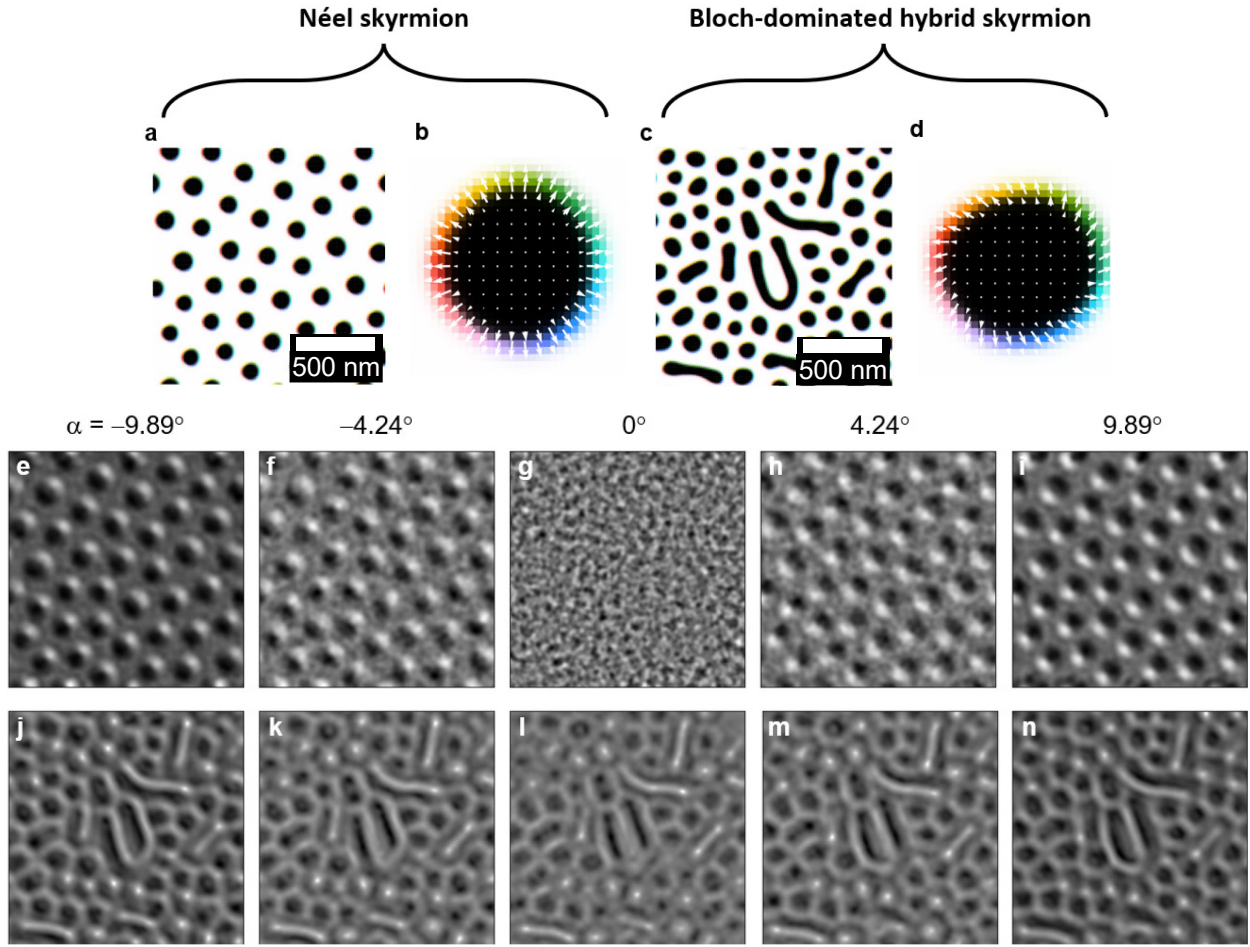

**Figure S14. Micromagnetic simulations and reconstructed LTEM images for Néel and Bloch-dominated hybrid skyrmions. a,** and **b,** Simulated Néel skyrmions and a magnified Néel skyrmion exposing its internal spin texture, respectively. **c,** and **d,** Simulated, Bloch-dominated spin texture, hybrid skyrmions and a magnified view of these hybrid skyrmion exposing their spin texture, respectively. **e,** and **i,** Reconstructed LTEM images for the simulated Néel skyrmions. A significantly reduced contrast is observed at the tilt angle $\alpha = 0°$, which is the characteristic of a Néel skyrmion. **j** and **n,** Reconstructed LTEM images from the simulated Bloch-dominated hybrid skyrmions.

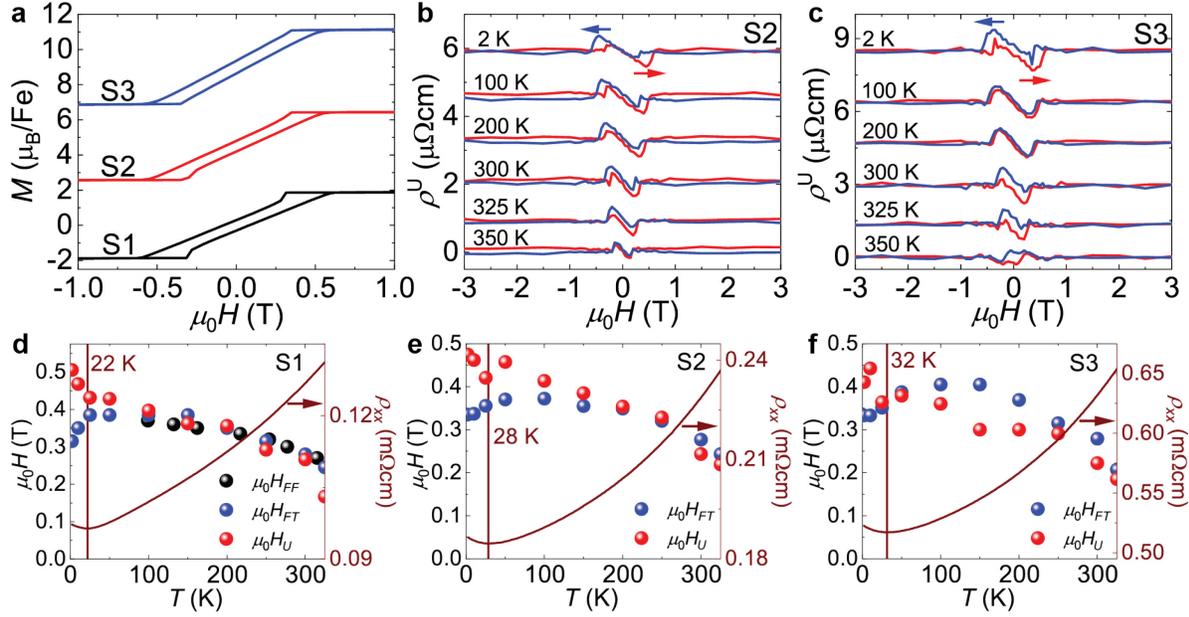

**Figure S15. Unconventional Hall effect and magnetic phase boundary in lower-quality samples.** Unconventional Hall effect for the high-quality sample (S1), whose data is included in the main manuscript, compared to the unconventional Hall signal collected from lower-quality samples (S2 and S3). For crystals S1, S2, and S3, the resistivity values at $T = 2$ K are 0.097, 0.187, and 0.523 mΩ cm, and those of the anomalous Hall conductivity at the same $T$ under $\mu_0 H = 1$ T are 419, 328, and 70 $\Omega^{-1}\text{cm}^{-1}$, respectively. **a,** Magnetization of crystals S1, S2, and S3. Traces are vertically displaced for the sake of clarity. The sudden magnetization step observed upon leaving the saturation plateau, weakens considerably as the sample quality decreases more clearly exposing the unconventional contribution to the Hall effect. **b and c,** Unconventional Hall effects collected from crystals S2 and S3, respectively. **d, e, and f,** Critical fields defining the boundary between FM and FiM ($\mu_0 H_{FF}$ and $\mu_0 H_{FT}$) states, as well as the peak position ($\mu_0 H_U$) observed in $\rho^U$ as functions of $T$ for the high-quality sample S1, and the lower-quality samples S2 and S3,

respectively. The brown solid line corresponds to the resistivity $\rho_{xx}$ of each sample as a function of $T$.

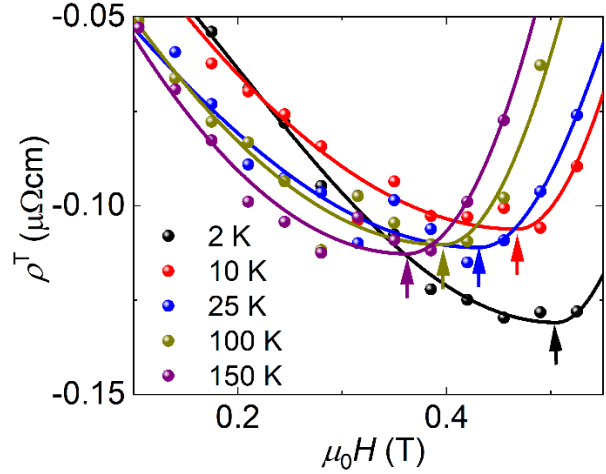

**Figure S16. Procedure used to extract the position of the maxima in the topological Hall signal.** Arrows indicate peak positions associated with the maxima in the topological Hall signal collected at distinct temperatures. We fitted the data to an asymmetric Gaussian function or BiGaussian function. The fitting lines are depicted as solid lines. The peak position moves to higher fields as the temperature is reduced, albeit it moves faster within the low temperature range (< 25 K).

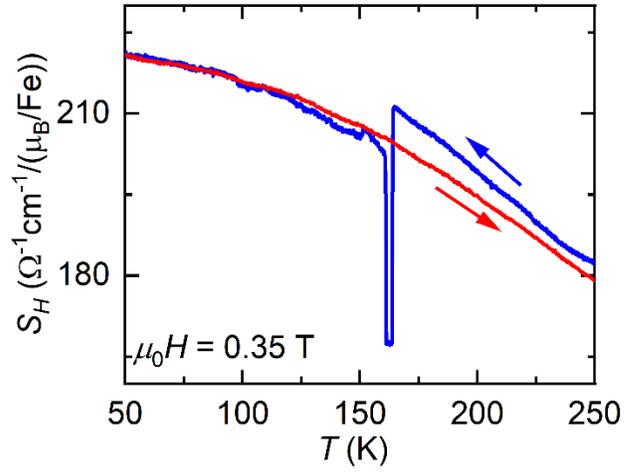

**Figure S17. $S_H$ as a function of the temperature upon cooling (blue line) and warming (red line) under $\mu_0 H$ = 0.35 T.** Difference in $S_H$ prior and after the magnetic phase-transition.

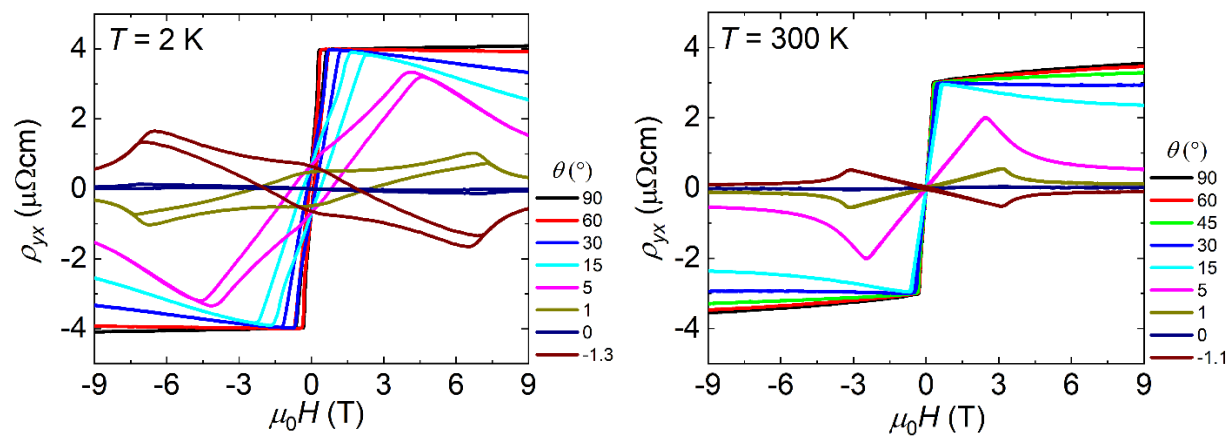

**Figure S18. Hall resistivity for a $Fe_3GaTe_2$ single crystal at $T = 2$ K (left) and $T = 300$ K (right) as a function of the magnetic field for several field orientations.** Corresponding experimental configuration is shown in Figure S8a.

**Table S1.** Crystallographic data, data collection, and refinement parameters (Single Crystal X-ray Diffraction).

| Formula | Fe$_{2.99}$GaTe2 |
|---|---|
| Space Group | *P6$_3$/mmc* |
| $a$ (Å) | 4.0793(3) |
| $c$ (Å) | 16.107(3) |
| $V$ (Å$^3$) | 232.12(5) |
| Z | 2 |
| Temperature (K) | 298 |
| $\theta$ (°) | 5.1-30.5 |
| $\mu$ (mm$^{-1}$) | 27.02 |
| Measured Reflections | 8229 |
| Independent Reflections | 172 |
| R$_{int}$ | 0.053 |
| $\Delta\rho_{max}$, $\Delta\rho_{min}$ (e$^-$/Å$^3$) | 1.68, -2.23 |
| $R_1$ [$F^2 > 2\sigma(F^2)$] | 0.027 |
| $wR_2$ ($F^2$) | 0.065 |

$R_1 = \Sigma ||F_o| - |F_c|| / \Sigma |F_o|$

$wR_2 = \{\Sigma[w(F_o^2 - F_c^2)^2] / \Sigma[w(F_o^2)^2]\}^{1/2}$

**Table S2. Fractional atomic coordinates and displacement parameters (Single Crystal X-ray Diffraction).**

| Site | Wyckoff | x | y | z | $U_{eq}$ (Å$^2$) | Occupancy |
|------|---------|---|---|---|------------------|-----------|
| Fe1  | 4e      | 0 | 0 | 0.67280(10) | 0.0114(4) | 1 |
| Fe2  | 2c      | ⅔ | ⅓ | ¾ | 0.0102(8) | 0.901(13) |
| Fe3  | 2a      | 0 | 0 | ½ | 0.0102(8) | 0.088(8) |
| Ga1  | 2d      | ⅓ | ⅔ | ¾ | 0.0215(5) | 1 |
| Te1  | 4f      | ⅔ | ⅓ | 0.59079(5) | 0.0125(3) | 1 |

An EADP command was added to constrain the atomic displacement parameter of the Fe2 and Fe3 sites to improve the accuracy of the freely refined occupancies.

**Table S3. Comparison between lattice parameters and Fe2/Fe3 site occupancies for different Fe$_3$GaTe$_2$ samples.**

| Crystal | $a$ (Å) | $c$ (Å) | $V$ (Å$^3$) | Fe(2) Occ. | Fe(3) Occ. |
|---------|-----------|------------|-------------|------------|------------|
| 1 | 4.0793(3) | 16.107(3) | 232.12(5) | 0.901(13) | 0.088(8) |
| 2 | 4.0780(6) | 16.095(4) | 231.80(9) | 0.87(2) | 0.117(13) |
| 3 | 4.0846(11) | 16.134(6) | 233.12(15) | 0.88(2) | 0.079(15) |
| 4 | 4.0732(7) | 16.115(7) | 231.55(13) | 0.88(2) | 0.079(13) |
| 5 | 4.0916(13) | 16.168(9) | 234.41(19) | 0.90(2) | 0.069(16) |

**Table S4. Stoichiometry of a crystal according to EDS.**

| #1 | Fe (%) | Te (%) | Ga (%) |
|---|---|---|---|
| P1 | 51 | 34.6 | 14.4 |
| P2 | 51.1 | 32.2 | 16.9 |
| P3 | 51.4 | 33.2 | 15.4 |
| P4 | 50.4 | 34.3 | 15.3 |
| P5 | 49.7 | 32.3 | 18 |
| Average | 50.7 | 33.3 | 16 |
| Std. dev. | 0.605 | 0.991 | 1.28 |

**Movie S1. LTEM images on cooling.** The angle between the normal direction of the sample surface and the magnetic field of $\mu_0 H = 250$ Oe was fixed at $-21°$. The temperature was cooled from $T = 373$ K to 123 K.

**Movie S2. LTEM images on warming.** The angle between the normal direction of the sample surface and the magnetic field of $\mu_0 H = 250$ Oe was fixed at $-21°$. The temperature was warmed from $T = 120$ K to 353 K.